\documentclass[prb,aps,showpacs,twocolumn]{revtex4}
\usepackage{graphicx}
\usepackage{epstopdf}
\usepackage{dcolumn}
\usepackage{bm}
\usepackage{color}
\usepackage{ulem}
\usepackage{amsfonts,amsthm}
\usepackage{amsmath}
\usepackage{amssymb}
\begin{document}
\newcommand{\bR}{\mbox{\boldmath $R$}}
\newcommand{\tr}[1]{\textcolor{black}{#1}}
\newcommand{\tb}[1]{\textcolor{black}{#1}}
\newcommand{\tm}[1]{\textcolor{black}{#1}}
\newcommand{\tg}[1]{\textcolor{black}{#1}}
\newcommand{\tc}[1]{\textcolor{black}{#1}}

\newcommand{\Ha}{\mathcal{H}}
\newcommand{\mh}{\mathsf{h}}
\newcommand{\mA}{\mathsf{A}}
\newcommand{\mB}{\mathsf{B}}
\newcommand{\mC}{\mathsf{C}}
\newcommand{\mS}{\mathsf{S}}
\newcommand{\mU}{\mathsf{U}}
\newcommand{\mX}{\mathsf{X}}
\newcommand{\sP}{\mathcal{P}}
\newcommand{\sL}{\mathcal{L}}
\newcommand{\sO}{\mathcal{O}}
\newcommand{\la}{\langle}
\newcommand{\ra}{\rangle}
\newcommand{\ga}{\alpha}
\newcommand{\gb}{\beta}
\newcommand{\gc}{\gamma}
\newcommand{\gs}{\sigma}
\newcommand{\vk}{{\bm{k}}}
\newcommand{\vq}{{\bm{q}}}
\newcommand{\vR}{{\bm{R}}}
\newcommand{\vQ}{{\bm{Q}}}
\newcommand{\vga}{{\bm{\alpha}}}
\newcommand{\vgc}{{\bm{\gamma}}}
\newcommand{\mb}[1]{\mathbf{#1}}
\def\vec#1{\boldsymbol #1}
\arraycolsep=0.0em
\newcommand{\Ns}{N_{\text{s}}}

\title{
Origin of high-$T_{c}$ superconductivity in doped Hubbard models and their extensions: \\
{Roles of uniform charge fluctuations}
}

\author{Takahiro Misawa and Masatoshi Imada}

\affiliation{%
Department of Applied Physics, University of Tokyo,
7-3-1 Hongo, Bunkyo-ku, Tokyo, 113-8656, Japan}

\date{\today}
\begin{abstract}
Doped Hubbard model is a 
simple model for the high-$T_{c}$ cuprate superconductors, while its ground state remains a challenge.
Here, by performing state-of-the-art variational Monte Carlo calculations 
for the strong-coupling Hubbard model, 
we find evidences that the $d$-wave superconducting phase emerges {always} 
near the phase separation region and the superconducting order has one-to-one 
correspondence with the enhancement of charge compressibility. The order as well as the phase 
separation are vulnerable to realistic intersite Coulomb 
interaction while the superexchange interaction enhances both. 
An appropriate combination of these two widens the stable superconducting phase. 
\end{abstract}

\maketitle

\section{Introduction}
The discovery of high-$T_{c}$ 
superconductivity in copper oxides{~\cite{Bednorz}}
triggers studies of  
the superconductivity 
induced by the strong electronic correlations.
After an enormous number of studies, 
the intrinsic phase diagram of the copper oxides is still not a completely resolved issue.
Most of the superconducting copper oxides have the dome structure of the critical 
temperature $T_c$ as a function of the hole doping concentration $\delta$ 
centered at the optimum value $\sim 0.15$,  after the quick disappearance of 
the antiferromagnetic order upon the doping to the Mott insulator of 
the mother materials.  

However, the multi-layer compound shows a wide coexistence 
region of the superconductivity and the antiferromagnetic order~\cite{Mukuda}. 
Recently the interface of La$_2$CuO$_4$/La$_{2-x}$Sr$_x$CuO$_4$, which is expected to 
realize purely two-dimensional superconductivity, {has strikingly shown} a pinning of $T_c$ at 
a {constant} value $\sim 40$K~\cite{WuBozovic} in marked contrast with the 
dome structure in bulk, which supports that the intrinsic nature of 
the copper oxides is described by an extended region of the phase separation (PS), if the long-ranged 
Coulomb interaction is screened by the interlayer screening. {At the interface, the phase separation {may occur between layers}.} 
The intrinsic phase diagram of the copper oxides without 
impurity and long-ranged Coulomb effects is still an actively debated issue. 

One of the most fundamental models to describe the high-$T_{c}$
superconductivity 
is the Hubbard model on the square lattice,
which only considers the nearest-neighbor hopping $t$ and
on-site Coulomb repulsion $U$ of electrons \tr{(details are shown in Sec.~\ref{sec:Model})}.
A large number of theoretical works including analytical and numerical calculations 
have been 
devoted {to} the Hubbard model 
~\cite{Furukawa1992,AimiGBMC,Yokoyama2012,Maier,Aichhorn,Khatami,Capone,Sordi,Gull,Chen_2013,MoriyaUeda,Eichenberger,CPMC} 
(Detailed comparison of previous studies are shown in \tr{Appendix~\ref{sec:comparison}}).
Many of works suggest that the superconductivity appears 
{near} half band filling for sufficiently large 
$U/t${~\cite{Eichenberger,Yokoyama2012,Maier,Aichhorn,Khatami,Capone,Sordi,Gull,Chen_2013,MoriyaUeda,FRGRMP}}.
However, numerically exact or high-precision calculations~\cite{Furukawa1992,AimiGBMC,CPMC}
do not {necessarily} show clear evidence of the high-$T_{c}$ superconductivity. 
Thus, the relation between strong electronic correlations 
and the high-$T_{c}$ superconductivity
still remains an unresolved issue {although there are many
proposals for origin of the high-$T_{c}$ superconductivity
~\cite{MoriyaUeda,FRGRMP,LeeRMP,EmeryKivelson,Chen_2013,Kivelson,Furukawa1992,Aichhorn,Capone,Chang,Sorella,Neuscamman,Stripe_tJ,
Tocchio,ImadaMQCP,ImadaSuper}.}
\tr{The Hubbard model tremendously simplifies the real materials. However, the 
prolonged controversy implies the significance of clarifying the superconductivity 
in the doped Hubbard models to understand the fundamental origin of the high-$T_{\rm c}$  
superconductivity provided that {reliable} theoretical calculations are performed.}

In this paper, \tr{by performing state-of-the-art calculations},
we show a direct and quantitative one-to-one correspondence 
between superconductivity and enhanced uniform charge susceptibility,
which clearly shows that the tendency for the PS is the origin of the $d$-wave superconductivity.
The present result also offers 
an intriguing implication to the recent interface experiment~\cite{WuBozovic}.
We further reveal roles of intersite Coulomb repulsion 
$V$ that reduces both superconducting phase and uniform charge fluctuations 
as well as roles of superexchange interaction $J$ that enhances both of them.

\section{ Model, method, and definitions of physical quantities}
\label{sec:Model}
We employ the standard Hubbard model on the square lattice, {defined by the Hamiltonian}
\begin{align*}
H=-t\sum_{\langle i,j\rangle,\sigma}(c_{i\sigma}^{\dagger}c_{j\sigma}+{\rm h.c.})
+U\sum_{i}n_{i\uparrow}n_{i\downarrow},
\end{align*}
where $c_{i\sigma}^{\dagger}$ ($c_{i\sigma}$) is the creation (annihilation)
operator on the $i$-th site with spin $\sigma$ and 
$n_{i\sigma}=c_{i\sigma}^{\dagger}c_{i\sigma}$ is the number operator.
The transfer integral $t$ is only taken for nearest-neighbor sites.
We take $N_{\rm s}=L\times L$ sites with 
periodic-periodic (PP) and 
antiperiodic-periodic (AP)
boundary conditions. We define the doping rate $\delta$ as
$\delta=1-N_{e}/N_{\rm s}$,
where $N_{e}=\sum_{i,\sigma}n_{i\sigma}$.
We add the off-site Coulomb and superexchange interactions defined as
\begin{align*}
H_{V}&=V\sum_{\langle i,j\rangle}n_{i}n_{j},\\
H_{J}&=J\sum_{\langle i,j\rangle}\boldsymbol{S}_{i}\cdot\boldsymbol{S}_{j}, 
\end{align*}
where $\bm{S}_{i}=1/2\sum_{\sigma,\sigma^{\prime}}
c^{\dagger}_{i,\sigma}\boldsymbol{\sigma}_{\sigma,\sigma^{\prime}}c_{i,\sigma^{\prime}}$
and $n_{i}=n_{i\uparrow}+n_{i\downarrow}$.

To study the ground-state of the doped Hubbard model,
we employ a many-variable variational Monte Carlo (mVMC)  method
combined with the quantum-number projection.
Our variational wave function is defined as
\begin{equation}
|\psi\ra =\sP_{\rm G}\sP_{\rm J}\sP_{\rm d-h}^{\rm ex}\sL^{K=0}\sL^{S=0}|\phi_{\rm pair}\ra,
\label{Eq:WF}
\end{equation}
where $\sP_{\rm G}$, $\sP_{\rm J}$, $\sP_{\rm d-h}^{\rm ex}$ 
are the 
{Gutzwiller~\cite{Gutzwiller}, 
Jastrow~\cite{Jastrow,CapelloJastrow}, 
and doublon-holon correlation factors~\cite{YokoyamaDH},} 
respectively~\cite{TaharaVMC_Full}.
The Gutzwiller factor punishes the double occupation of electrons on 
the same site through the variational parameters $g$ defined as
\begin{align*}
  \sP_{\text{G}} = \exp(-g\sum_{i}n_{i\uparrow} n_{i\downarrow}).
\end{align*}
The Jastrow factors are defined as  
\begin{align*}
  \sP_{\text{J}} = \exp(-\frac{1}{2} \sum_{i,j} v_{ij} n_{i} n_{j}),
\end{align*}
where the long-range part drives the distinction between the metal and insulator{~\cite{CapelloJastrow}}. 
The doublon-holon correlation factors~\cite{YokoyamaDH} are defined as
\begin{align*}
 \sP_{\text{d-h}}^{\text{ex}} =& \exp\biggl[ - \sum_{m=0}^{2}\sum_{\ell=1,2} \ga_{(m)}^{(\ell)} \sum_{i} \xi_{i(m)}^{(\ell)} \biggr], 
\end{align*}
where  $\xi_{i(m)}^{(\ell)}$ is a many-body operator which is diagonal in the real-space representations. 
When a doublon (holon) exists at the $i$-th site and $m$ holons (doublons) surround at the $\ell$-th nearest neighbor, $\xi_{i(m)}^{(\ell)}$ gives $1$. Otherwise, $\xi_{i(m)}^{(\ell)}$ gives $0$. 
The spin (momentum) quantum number projection operator 
$\sL^{S=0}$ ($\sL^{K=0}$)
restores $SU$(2) spin symmetry (translational symmetry) with
the total spin $S=0$ (total momentum $K=0$). 
{These projections substantially improve the accuracy of cluster properties, make the size 
dependence smaller and the extrapolation to the thermodynamic limit easier~\cite{TaharaVMC_Full}}.

The one-body part $|\phi_{\rm pair}\ra$ is
the generalized pairing wave function defined as
\begin{align}
|\phi_{\rm pair}\ra= \Big[\sum_{i,j=1}^{\Ns}f_{ij}c_{i\uparrow}^{\dag}c_{j\downarrow}^{\dag}\Big]^{N_{e}/2} |0 \ra,
\label{Eq:onebody}
\end{align}
where $f_{ij}$ denotes the variational parameters~
{(Details of $f_{ij}$, see Refs.~\onlinecite{gros1989physics,BajdichPRB,TaharaVMC_Full}).} 
In this study, we
allow $f_{ij}$ to have $2\times2$ sublattice structure
or equivalently we have $2\times2\times N_{\rm s}$ 
independent variational parameters for one-body part.
All the variational parameters are simultaneously
optimized by using the stochastic 
reconfiguration method~\cite{Sorella_PRB2001,TaharaVMC_Full}.
The variational function $|\psi\ra$
in Eq.~(\ref{Eq:WF})
can flexibly describe paramagnetic metals, 
the antiferromagnetic phase, and superconducting phases
as well as their fluctuations and/or coexistence.
It is important to fully optimize the long-range part of $f_{ij}$ to realize 
states with strong fluctuations and well-developed short-ranged order as well as strongly renormalized metals as we detail later.
Actually, by extending the {2$\times$2} sublattice structures of the variational 
parameters $f_{ij}$, we confirmed that the accuracy of the energy is 
improved. 

\begin{figure}[t!]
	\begin{center}
		\includegraphics[width=7cm,clip]{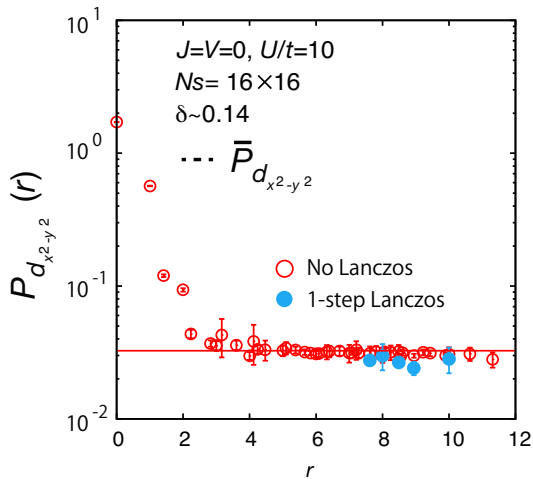}   
	\end{center}
\caption{Distance ($r$) dependence of superconducting correlation for $\delta\sim0.14$, $J=V=0$, and $U/t=10$.
The system size is $N_{\rm s}=16\times16$ and 
AP boundary condition is used.
Results of first-step Lanczos method are shown by (light-blue) closed circles. }
\label{fig:SCNe110}
\end{figure}

Furthermore, by applying the power Lanczos method~{\cite{PowerLanczos}}, we can 
also substantially improve the energy. 
In the $N$-th step power Lanczos method, we multiply 
Hamiltonian to the variational wavefunctions as follows:
\begin{align}
|\psi_{n}\ra = \Big(1+\sum_{n=1}^{N}\alpha_{n}H^{n}\Big)|\psi\ra,
\end{align}
where $\alpha_{n}$ are the variational parameters. By choosing 
$\alpha_{n}$ to lower the energy, we can systematically improve the variational wave functions, 
\tr{as we see later in Fig.~\ref{fig:var_ext}}.
However, through the careful examination of 
such extensions, we confirmed that estimates of 
the physical properties 
(superconducting correlations, antiferromagnetic correlations, etc .) change little 
(for example, see Fig.~\ref{fig:SCNe110}). 
In addition, numerical cost of such extensions is demanding.
Therefore, to perform the comprehensive calculations for the doped 
Hubbard with additional intersite interactions, 
we have used the present tractable variational wave functions.
Nevertheless, we again emphasize that the estimates of the physical 
properties themselves are accurate enough and our conclusions do not change.

To discuss the condensation energy,
we generate two different wave functions, i.e., 
normal and superconducting wave functions by choosing proper initial states.
We employ the non-interacting Fermi sea for {the} normal state, and
BCS $d$-wave superconductivity state for 
superconducting phase {as the initial states}~\cite{TaharaVMC_Full}. 
By optimizing these initial states, we obtain normal 
and superconducting states.
In the strong coupling region, the antiferromagnetic order 
appears near half filling as the normal state as a result of the optimization, although we do not assume the 
antiferromagnetic order as an initial state, which means that the paramagnetic normal state is unstable.

To determine the ground state of the doped Hubbard model,
we calculate spin structure factor and equal-time superconducting correlation.
The spin structure factor is defined as 
\begin{align*}
S(\bm{q})=\frac{1}{3N_{s}}\sum_{i,j}
\langle \boldsymbol{S}_{i}\cdot\boldsymbol{S}_{j}\rangle e^{i\bm{q}\cdot(\bm{r}_{i}-\bm{r}_{j})},
\end{align*}
and the equal-time superconducting correlations 
are defined as
\begin{align*}
P_{\alpha}(\bm{r})=\frac{1}{2N_{\rm s}}\sum_{\bm{r}_{i}}
\langle\Delta_{\alpha}^{\dag}(\bm{r}_{i})\Delta_{\alpha}(\bm{r}_i+\bm{r})+\Delta_{\alpha}(\bm{r}_i)\Delta_{\alpha}^{\dag}(\bm{r}_i+\bm{r})\rangle.
\end{align*}
{In actual calculations, to reduce numerical cost, we restrict summation with respect $\bm{r}_{i}$
to $\bm{r}_{i}=0$.}
Superconducting order parameters  $\Delta_{\alpha}(\bm{r}_{i})$ {are} defined as
\begin{align*}
\Delta_{\alpha}(\bm{r}_i)=\frac{1}{\sqrt{2}}
\sum_{\bm{r}}f_{\alpha}(\bm{r})({c}_{\bm{r}_i\uparrow}{c}_{\bm{r}_i+\bm{r}\downarrow}-{c}_{\bm{r}_i\downarrow}{c}_{\bm{r}_i+\bm{r}\uparrow}).
\end{align*}
Here, $f_{\alpha}(\bm{r})$ is the form factor that
describes the symmetry of the superconductivity.
For $d_{x^2-y^2}$ superconductivity, we define
\begin{align*}
f_{d_{x^2-y^2}}(\bm{r})&=\delta_{r_{y},0}(\delta_{r_{x},1}
+\delta_{r_{x},-1})-\delta_{r_{x},0}(\delta_{r_{y},1}+\delta_{r_{y},-1}) ,  
\end{align*}
where $\delta_{i,j}$ denotes the Kronecker's delta and $\bm{r}=(r_{x},r_{y})$.
We define long-range average of the superconducting correlation as
\begin{align*}
\bar{P}_{d_{x^2-y^2}}&=\frac{1}{M}\sum_{2<r=|\bm{r}|<L-1}P_{d_{x^2-y^2}}(\bm{r}),
\end{align*}
where $M$ is the number of vectors satisfying $2<r<L-1$.
{As shown in Fig~.~\ref{fig:SCNe110}}, the criterion $r>2$ is, within 
the present purpose, practically a sufficient probe to see whether 
the pairing order-parameter correlation is saturated 
to a nonzero value and offers a good measure for 
the square of the order parameter in the long-range ordered superconducting state.
{We also note that the first-step power Lanczos method does not {essentially} change
the superconducting correlations {as we see in Fig.~\ref{fig:SCNe110}}}.

We also calculate the chemical potential 
by using the relation
\begin{align}
\mu(\bar{N})=\{E(N_{1})-E(N_{2})\}/\{N_{1}-N_{2}\}-{\frac{U}{2}},
\label{Eq:mu}
\end{align}
$E(N_{1})$ is the total energy at filling $N_{1}$ and
$\bar{N}=(N_{1}+N_{2})/2$. 
To directly compare with previous calculations~\cite{Furukawa1992,Furukawa93},
we subtract constant value $U/2$.
To reduce the finite-size effects, we perform calculation only at
the electron densities that satisfy the closed-shell 
condition in the non-interacting case~\cite{Furukawa1992,Furukawa93}.

The nonzero condensation energy $\Delta E= (E_{\rm SC}-E_{\rm Normal})/N_s$ is 
defined when the superconducting (with energy $E_{\rm SC}$) and normal states ($E_{\rm Normal}$)
exist as local minima.  The normal state is not necessarily the paramagnetic 
state but can be another symmetry broken state such as the 
antiferromagnetically ordered state, if it has a lower energy 
than the paramagnetic state.  It is remarkable that in the present calculation, 
if the superconducting state with a nonzero order parameter exists, 
it always has a normal state as local minima as well. 
The transition from the normal to the superconducting states by 
reducing the doping concentration from the overdoped region is always a weak 
first-order transition where the superconducting order parameter jumps from 
zero to a small nonzero value in the ground state. 
For instance, \tr{as we show later}, at $(V/t=0, J/t=0)$, $(V/t=0,J/t=0.5)$, $(V/t=1,J/t=0.5)$ and $(V/t=2,J/t=0.5)$, 
the superconducting state emerges as a metastable state at $\delta \sim  0.25, 0.33, 0.29, 0.28$ 
while it becomes the ground state only {for} $\delta {\lesssim}  0.22, 0.31, 0.27, 0.28$, respectively. 
The first-order jump decreases with the increase in $V/t$ suggesting an existence of the 
tricritical point at around  $(V/t=2,J/t=0.5)$. Toward  half filling, the order parameter of 
the superconducting state looks continuously {going} to zero, which is 
connected to the antiferromagnetic Mott insulator. 
Here, again the non-superconducting state continues to exist as a metastable excited state.

In connection with the experimental measurement of the 
condensation energy by the specific heat or the upper critical field, 
the present definition is not exactly identical each other because 
the normal state in the experiment usually excludes the 
magnetic order as the normal state, for instance. This means that the 
experimental value overestimates the true condensation energy. However, the present 
definition certainly gives more useful criterion to 
determine whether the superconducting state is the true ground state or not.

The normal state is defined as the state that has vanishing superconducting order 
within the numerical accuracy.  It does not exclude the possibility of a state with a 
tiny order parameter expected from the Kohn-Luttinger mechanism~{\cite{KohnLuttinger}}. 
In addition, the normal state we obtained has a robust and 
developed superconducting correlation with the extended $s$-wave 
order parameter with the form factor $\cos k_x +\cos k_y$,  which scales to 
zero in the thermodynamic limit within the numerical accuracy. 

Monte Carlo sampling of real space configurations of the electrons is
employed to calculate physical quantities following the standard procedure~\cite{TaharaVMC_Full}.
The number of Monte Carlo samples for the calculation of physical quantities is
typically 128 000. The statistical error of the Monte Carlo sampling estimated from
a number of independent computations is indicated
in the last parentheses in the numerical data as well as error bars in the plots in figures.

\section{ Results }
\subsection{ Simple Hubbard model ($V=J=0$) }
To examine the origin of 
high-$T_{c}$ superconductivity
in the Hubbard model, 
we employ mVMC 
method~\cite{TaharaVMC_Full} (for validity of the method, \tr{see Appendix~\ref{sec:benchmark}}).
This method enables us to perform high-precision 
calculations 
{under spatial and temporal fluctuations of spin and charge on equal footings with a sufficient flexibility of wavefunctions}, which are
important in strongly correlated systems.

\begin{figure}[ht!]
	\begin{center}
		\includegraphics[width=7cm,clip]{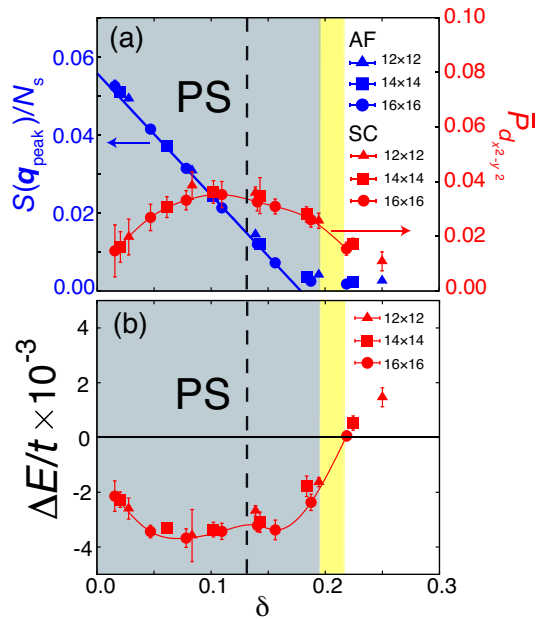}   
	\end{center}
\caption{{(color online).} (a)~Doping ($\delta$) dependence of 
averaged $d_{x^{2}-y^{2}}$-wave 
superconducting correlations $\bar{P}_{x^2-y^2}$ 
and peak values of spin structure factors  $S(\bm{q}_{\rm peak})$
{for $U/t=10$ and $V=J=0$}.
{Doping rate $\delta$ is defined as $\delta=1-N_{e}/N_{\rm s}$, where
$N_{e}$ ($N_{\rm s}$) represents number of electrons (system size).}
We note that the incommensurate spin orders or stripe phases 
are not found
in the relevant doping region $\delta\lesssim0.2$ even when we employ
large sublattice structures. 
{We also note that the charge structure factors have no significant peak at $q\neq0$.}
(b)~Doping dependence of condensation energy $\Delta E$. 
The condensation energy is defined as $\Delta E=(E_{\rm SC}-E_{\rm Normal})/N_{\rm s}$, where 
$E_{\rm SC}$ ($E_{\rm Normal}$) is the total energy of the superconducting phase (normal phase).
The calculations are performed for {sizes of} $N_{\rm s}=12\times12,~14\times14, 16\times16$ {on the square lattice}, and
we confirm that the finite-size effects are negligibly small. The shaded region denotes the PS region and 
the black dashed line represents the spinodal point. Details of PS are shown
in {the} main text and  Fig.~\ref{fig:muN}. The superconducting phase without PS remains only in the yellow region.
In the present plots and the plots in the later figures,
the error bars indicate the estimated statistical errors of 
the Monte Carlo sampling.
}
\label{fig:U10}
\end{figure}%

\begin{figure}[htp!]
	\begin{center}
		\includegraphics[width=7cm,clip]{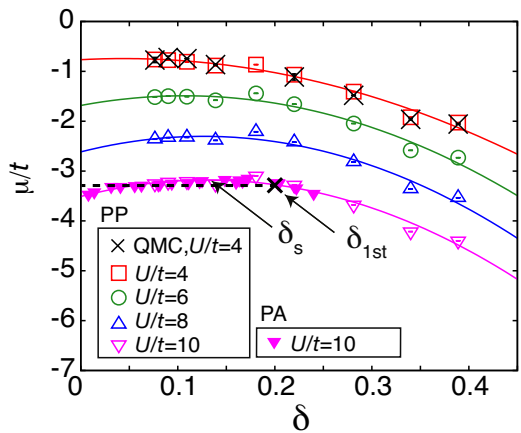}   
	\end{center}
\caption{{(color online).} Doping dependence of chemical potential for $U/t=4,8,6,10$,
{$V=J=0$, and system sizes $L=6,8,10,12,14,16$, where $N_{\rm s}=L\times L$.} 
{We note that different size results are essentially on the same curve.}
{For $U/t=4$,}
our mVMC successfully reproduces the results of quantum Monte 
Carlo (QMC) represented by black crosses~\cite{Furukawa1992}.
By fitting the chemical potential with the second-order polynomials,
we estimate the spinodal point, {where {$(dn/d\mu)^{-1}=0$}}. 
We also estimate the PS region ($\delta<\delta_{\rm 1st}$) 
by {performing Maxwell's construction \tg{using the fitted} second-order polynomials.}
{
Maxwell's construction for $U/t=10$ is shown by (black) dotted line.}
For $U/t=10$, we estimate that the PS occurs for $\delta<\delta_{\rm 1st}\sim0.195$.
We also estimate that the spinodal point, in which the 
charge compressibility diverges ($\chi_{c}^{-1}=0$),
is located at $\delta_{\rm s}\sim0.178$ for $U/t=10$.
{To ensure the existence of the PS, we further perform
the {first-step power Lanczos calculations}~(see Fig.~\ref{fig:1stLz} in Appendix F)
and 
{confirm that Lanczos step changes $\mu$ little.}
This result indicates that improvement of energy 
affects 
the PS \tm{region little}.}}
\label{fig:muN}
\end{figure}%

Figure \ref{fig:U10} shows the doping dependence of several physical properties for $U/t=10$;
peak value of the spin structure factor $S(\bm{q}_{\rm peak})/N_{\rm s}$,
which 
\tg{is the} square of the antiferromagnetic ordered moment, 
and average value of 
superconducting correlation $\bar{P}_{d_{x^2-y^2}}$
at long distance with the $d_{x^{2}-y^{2}}$ symmetry,
{corresponding to the square of the superconducting order parameter.} 
We also plot the condensation energy {$\Delta E$.}

 We find the $d_{x^{2}-y^{2}}$-wave superconducting phase 
{only} in the strong coupling region {$U/t\gtrsim6$}, 
which is consistent with previous studies\cite{Furukawa1992,AimiGBMC,Eichenberger,Yokoyama2012,Aichhorn,Capone,Sordi}.
{For instance, at 
$U/t=10$,
the $d$-wave superconductivity emerges for} $\delta\lesssim 0.2$ as shown in Fig.~\ref{fig:muN}. 
Both $\Delta E$ and $\bar{P}_{d_{x^2-y^2}}$
have dome structures around $\delta\sim 0.1$.
The {antiferromagnetic quantum critical point (AFQCP)} 
where the antiferromagnetic spin fluctuations diverge,
appears at $\delta\sim0.18$.
The $d$-wave superconductivity coexists with the antiferromagnetism in the 
ground state for $\delta\lesssim0.18$. 
The coexistence has been theoretically studied before in 
{several} different contexts~{\cite{Aichhorn,Capone,tJAFSC_PRB1990,Giamarchi_PRB1991}}. 
{The coexistence is basically consistent with the multilayer cuprates~\cite{Mukuda},
 where the PS may be suppressed by the interlayer self-doping.}


To examine the effects of charge fluctuations,
the doping dependence of
the chemical potential $\mu$ 
(\tr{see Appendix~\ref{sec:Nq}} for the charge structure factor in PS region) is shown in Fig.~\ref{fig:muN},
where the uniform
charge susceptibility $\chi_{c}\equiv dn/d\mu$ 
monitors the charge fluctuation.
{The} spinodal point of doping ($\delta_{\rm s}$), where charge fluctuations diverge ($\chi_{c}^{-1}=0$)
is found to increase 
{at} larger $U$.
{Accordingly, the} PS region 
becomes wider by increasing $U/t$. 
If we enforce the charge uniformity,
superconducting correlation has the maximum 
{around} $\delta_{\rm s}\sim0.14$ ({the spinodal point depicted by} dashed black line in Fig.~\ref{fig:U10}), for $U/t=10$.
This indicates that the enhanced charge fluctuations 
stabilize the superconducting phase around half filling.

{However, if the long-range Coulomb interaction 
is suppressed as in the Hubbard model, the present result indicates that in a wide region of 
the nominal doping concentration, the system undergoes a 
{real-space} PS into the antiferromagnetic Mott insulator 
and the superconducting region with the pinned $T_c$.  
This prediction is in striking agreement with 
the recent interfacial superconductivity~\cite{WuBozovic}.  }

\subsection{ Effects of inter-site interactions  }

\begin{figure}[tp!]
	\begin{center}
		\includegraphics[width=8cm,clip]{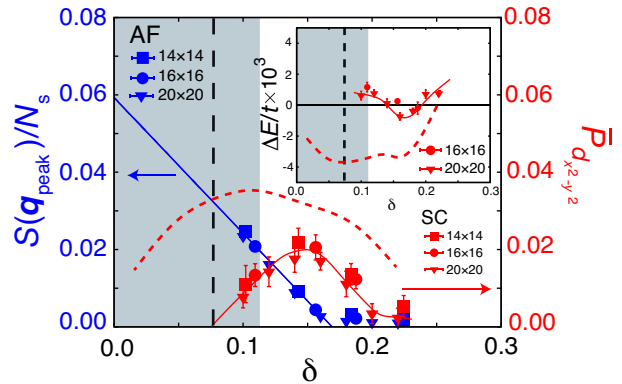}   
	\end{center}
\caption{{(color online).} (a)~Doping dependence of 
$\bar{P}_{x^2-y^2}$
and 
$S(\bm{q}_{\rm peak})$ at {$V/t=1$, $U/t=10$, and $J=0$} for
superconducting phase. 
In inset, condensation energy $\Delta E$ is plotted as a function of $\delta$.
{For comparison,
we plot results of $U/t=10$ and $V/t=0$ by broken lines}.
The calculation has been done up to 20$\times$20 lattices. {Notations are the same as Fig.~\ref{fig:U10}.}}
\label{fig:V1}
\end{figure}

Here, to control 
the charge fluctuations, 
we introduce nearest-neighbor 
Coulomb interactions $V$ (${H_{V}=V\sum_{\langle i,j\rangle}n_{i}n_{j}}$),
which 
indeed inevitably exit in real materials
~(\tr{see also Appendix \ref{sec:comparison} for previous studies}).
{As we see in {Fig.~\ref{fig:V1},}} 
although small $V/t=1$ drastically shrinks the PS region (gray shaded region),
the antiferromagnetic ordered moment and
the AFQCP does not change {appreciably}. 
Although the superconducting correlations have the peak around the AFQCP, 
{the condensation energy is largely reduced to almost zero as shown in { the inset of Fig.\ref{fig:V1}.}}
This result {supports} that the superconducting phase 
is predominantly  
stabilized by the enhanced charge fluctuations.
We note that the next-nearest hopping 
$t^{\prime}$ destabilizes 
the superconductivity in accordance with the shrinkage of the PS, 
which corroborates this {conclusion} \tr{(see also Appendix \ref{sec:tp})}.

It is also an intriguing issue to examine whether 
the instability toward the phase separation at the wavenumber $q=0$ 
can be converted into the instability toward charge ordering at nonzero $q$ 
observed in some cases of the cuprates 
by employing a realistic off-site Coulomb interactions.
\tr{In this calculation, we do not find any signatures of the 
charge ordering as shown in Fig.~\ref{fig:Nq} in Appendix \ref{sec:Nq}. }

To further understand the interplay of {spin fluctuations} and 
the instability toward the PS,
{by keeping $V=0$,}
we introduce the 
{nearest-neighbor} superexchange interactions $J$ 
{($H_{J}=J\sum_{\langle i,j\rangle}\boldsymbol{S}_{i}\cdot\boldsymbol{S}_{j}$)}
{that does not follow the standard relation $J_{\rm eff}\sim 4t^{2}/U$.} 
{In reality,} {$J$} can be induced by the $d$-$p$ 
hybridizations in cuprate superconductors 
beyond the single band framework~\cite{ZhangRice,Raimondi,Muller}.
\tr{
It has been repeatedly discussed in the literature that the superexchange 
interaction is derived from the three-band $d$-$p$ model for the cuprate 
superconductors in a nontrivial fashion without resorting to the single-band Hubbard model. 
Indeed the Zhang-Rice singlet~\cite{ZhangRice} produces the superexchange term $J$ that is rather 
independent of the expectation from the single band Hubbard model in the strong coupling limit. 
There exist several attempts to understand spin-dependent residual interactions 
within the single band description but beyond the Hubbard model with a finite $U$ but with an additional $J$~\cite{Raimondi,Muller}, 
while it is not well settled how the residual spin-dependent interaction should 
emerge quantitatively within the single band approach. 
In this circumstance, it is helpful and insightful to understand the role of 
residual superexchange-type interaction in the mechanism of 
superconductivity by taking the amplitude of $J$ as a parameter. 
}

{
As illustrated in Fig.~\ref{fig:J05VDep}(a), 
finite $J/t=0.5$ largely enhances the PS region, {while the antiferromagnetic order 
does not change appreciably.}
Accompanied by the 
enhanced 
charge fluctuations,
the condensation energy becomes an order of magnitude larger.
Because the AFQCP is close to the spinodal point as shown in
Fig.~\ref{fig:J05VDep}(b),
this {significantly} enhanced 
superconducting phase 
{may be understood from}
the synergetic effects of 
spin 
and charge fluctuations.
{We later emphasize the importance of {\it short-ranged} fluctuations.
However, anyway, this phase is again preempted by the PS.}

In addition to $J/t$, we again add 
$V$.
{As we see} in Fig.~\ref{fig:J05VDep}(a), 
by increasing $V/t$, locations of AFQCP do not change appreciably,
while locations of the spinodal point rapidly
approach half filling.
In connection with the {suppressed} charge fluctuations, 
the condensation energy is {significantly} reduced,
{again} suggesting {the key role of the proximity of the PS} 
in establishing high-$T_{c}$ superconductivity. 
{However, it is remarkable that, {for the coexisting $J$ and $V$}, the superconducting phase {with a substantial condensation energy} 
{survives} in a wide range ($0.1\lesssim\delta\lesssim0.3$ for $V/t=2$) outside
the PS region.}

\begin{figure}[tp!]
	\begin{center}
		\includegraphics[width=8cm,clip]{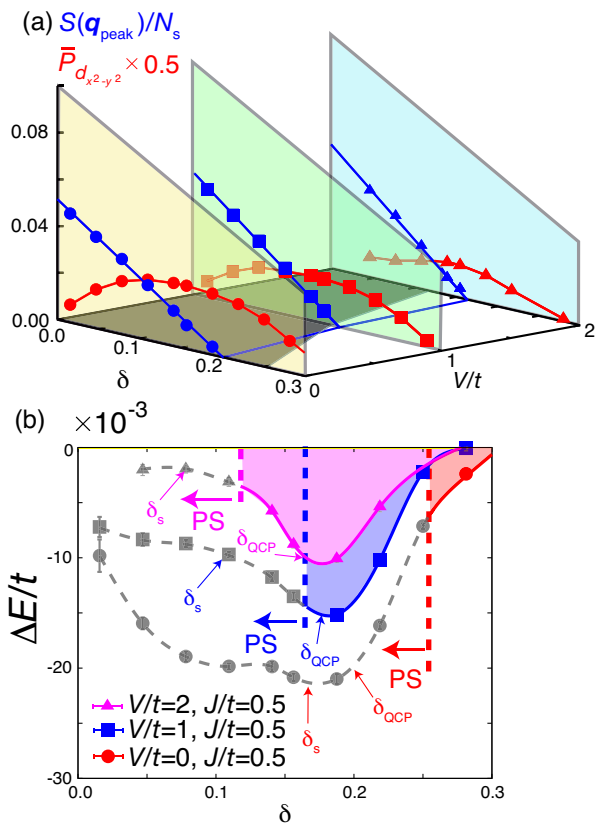}   
	\end{center}
\caption{{(color online).} (a) Doping dependence of 
$\bar{P}_{x^2-y^2}$
and 
$S(\bm{q}_{\rm peak})$ for $V/t=0, 1, 2$
{and fixed $U/t=10$ and $J/t=0.5$.}
{Additional $J$ significantly enhances the superconducting
correlations 
(see also  \tr{Fig.~\ref{fig:S}}).}
The shaded region {and the blue line in the bottom panel} 
represent the PS region and 
the position of the AFQCP, respectively.
(b) Condensation energy as a function of $\delta$.
In the PS region, condensation energy is plotted by gray symbols.
{The positions of the spinodal point ($\delta_{\rm s}$) and
the AFQCP ($\delta_{\rm QCP}$) are also plotted.} 
Solid and broken curves are guides for eyes.
}
\label{fig:J05VDep}
\end{figure}%

\begin{figure}[htbp!]
  \begin{center}
    \includegraphics[width=8cm,clip]{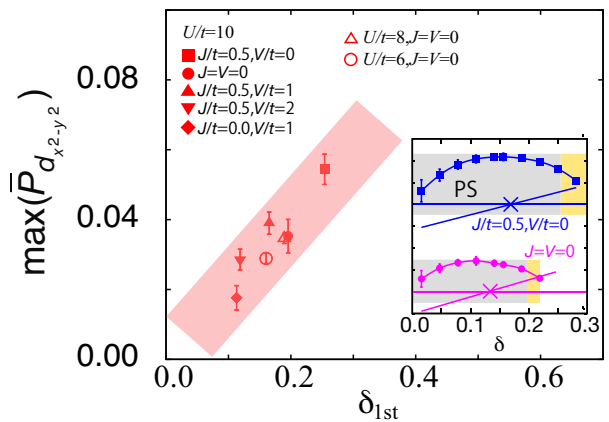}
  \end{center}
\caption{{(color online).} 
\tm{Relation between the peak value of the superconducting correlation 
${\rm max}(\bar{P}_{d_{x^{2}-y^{2}}})$ and the 
strength of enhancement of charge fluctuations characterized by
{width of the PS region $\delta_{\rm 1st}$.} 
Inset: Doping concentration dependence of ${20\times}\bar{P}_{d_{x^{2}-y^{2}}}$ 
(curves with symbols) and  ${0.1\times}\chi_c^{-1}$ {(lines passing crosses)} for two examples 
with offset in the ordinate for clarity. The crosses represent 
the spinodal point $\chi_c \rightarrow \infty$. 
\tg{The notation for the shaded zone is the same as Fig.~\ref{fig:U10}.} 
}
}
\label{fig:chiC}
\end{figure}%

The large condensation energy is ascribed mainly to two local sources: 
One is that 
{the double occupancy $D$} is largely 
reduced in the superconducting phase than that in the normal phase, 
which leads to the gain in the onsite Coulomb energy.
This is because, the $d$-wave pair prohibits the double occupation 
{strictly} by symmetry,
{which is particularly effective when $D$ remains {not small} in the normal phase
(as around $\delta\sim 0.1$)~(\tr{Fig.~\ref{fig:EkinEU}}  shows how the reduced $D$
in the superconducting state enhances $\Delta E$.)}
{This mechanism cannot be captured by the $t$-$J$ model.}
 {If $J>0$}, the other source is the 
antiferromagnetic correlation $\bm{S}_i\cdot\bm{S}_j$: 
{The {superconducting} order enhances the underlying nearest-neighbor ``antiferromagnetic" 
correlations even when $J=0$, which provides the energy 
gain immediately when {$J> 0$} \tr{(See Appendix \ref{sec:condene})}. The long-range part of antiferromagnetic correlation does not 
directly contribute to
this gain.}

{
The strong coupling nature of high-$T_{\rm c}$ superconductivity emerges {\it not} 
from the long-ranged part and the quantum criticality but rather from the local binding, 
as expected when approaching the regime of BEC. 
{This local attractive interaction
leads to Cooper pairing but does not necessarily
lead to PS. This is because the PS signaled by {the convex curve with a peak structure} in the
chemical potential as in Fig.~\ref{fig:muN}  is mainly caused by the contribution of the
kinetic energy part 
{in} the chemical potential, 
which is evidenced in \tr{Fig.~\ref{fig:muN_2} of Appendix \ref{sec:condene}}.
This peak in the kinetic energy is efficiently suppressed by $V$ rather
independently of the emergence of the local attractive interaction.}
{While $V$ suppresses PS,} {some choices} of $V$ and $J$ {largely} strengthen 
the energy gain from $D$ {because of the enhanced $D$ in the normal state.} 
This is the reason why {an} appropriate combination of $V$ and $J$ stabilizes 
the high-$T_{c}$ superconductivity without PS {in an extended region}.
}{It implies that the superconducting stability is not a 
universal property but largely relies on material details. It requires a 
reexamination of the conditions for the emergence of high-$T_{\rm c}$ superconductivity.  The necessity of both $V$ and $J$ also 
requires careful analyses how they are derived quantitatively from first principles.}

\tm{
To see relation between enhanced uniform charge fluctuations and 
the stability of superconductivity, 
we plot in Fig.~\ref{fig:chiC} the \tg{relation between} 
the maximum value of $\bar{P}_{d_{x^{2}-y^{2}}}$ \tg{and} 
{the width of the PS region}
in various cases of the superconducting state (main panel).
{Wider PS region indicates that charge fluctuation becomes more enhanced.}
The relation shows that a clear 
correlation between ${\rm max}(\bar{P}_{d_{x^{2}-y^{2}}})$ and 
{PS region}, 
indicating that the enhanced charge fluctuation stabilizes the superconductivity~\cite{ImadaMQCP}.    
In the inset,
we plot the doping {concentration} dependence of
the superconducting correlation and $\chi_c^{-1}$ for two typical examples. 
{In addition to the correspondence 
between ${\rm max}(\bar{P}_{d_{x^{2}-y^{2}}})$ and 
{PS region}, 
in all the cases we studied, 
the peaks of $\bar{P}_{d_{x^{2}-y^{2}}}$ are located at 
the concentrations close to the spinodal points (crosses), 
indicating again the one-to-one correspondence between $\bar{P}_{d_{x^{2}-y^{2}}}$ and the charge fluctuation.}
}


\section{Summary}
To summarize, the origin of the high-$T_{c}$ superconducting phase 
in the doped Hubbard model is found primarily as arising from the phase separation instability.
\tr{This conclusion suggests that the high-$T_{c}$ 
superconductivity is not necessarily a generic property of the doped Mott insulators,
but depends sensitively on the material specific parameters, particularly on the
intersite interactions, which gives a clue to
understand the strong material dependence of $T_c$.
Realistic intersite Coulomb repulsions $V$, which are often 
ignored in the literature, is by itself severely destructive to the 
superconductivity. However, it significantly contributes to widen the
high-$T_{\rm c}$ superconducting region 
without the phase separation,
if it is properly combined with the antiferromagnetic correlations such as superexchnage $J$.}

Controlling the charge fluctuation through the off-site 
interactions 
possibly by tuning \tr{a screening layer adjacent to conducting layer and the control of} 
dielectric constant offers a
possible way to stabilize the high-$T_{c}$ superconducting phase. 
Though it is not so easy,  
an interesting future issue is to find a way to suppress the ratio of
the off-site to on-site interactions by keeping a large on-site interaction in real materials
with the help of  {\it ab} {\it initio} calculations~\cite{ImadaMiyake}.
In this respect, the recently studied interfacial superconductivity~\cite{WuBozovic} offers
a promising way and supports
the relevance of the present phase diagram with an extended region of PS 
as the genuine one if the long-ranged Coulomb interaction
is screened on a single layer by the capacitor formation with the neighboring metallic layers.

\vspace{0.0cm}
\begin{acknowledgments}
The authors thank 
Daisuke Tahara and Satoshi Morita 
for providing them with efficient mVMC codes.
\tr{To compute the Pfaffian of skew-symmetric matrices, 
we employ the PFAPACK~\cite{PFAPACK}. A part of 
algorithms used in exact diagonalization is 
based on TITPACK version 2 coded by Hidetoshi Nishimori.}
{They also thank Antoine Georges for helpful comments.}
This work is financially supported by MEXT HPCI Strategic Programs 
for Innovative Research (SPIRE) and Computational Materials Science Initiative (CMSI).
Numerical calculation was partly carried out at the Supercomputer Center, 
Institute for Solid State Physics, Univ. of Tokyo. 
Numerical calculation was also partly carried out at K computer at 
RIKEN Advanced Institute for Computational Science (AICS)
{under grant number hp120043, hp120283 and \tm{hp130007}}. 
This work was also supported by Grant-in-Aid for 
Scientific Research {(No. 22104010, No. 22340090, and No. 23740261)} from MEXT, Japan.
\end{acknowledgments}

\appendix

\section{ Comparison with previous studies}
\label{sec:comparison}
\begin{table*}[hbt!]
  \begin{flushleft}
  \scalebox{1}{
  \begin{tabular}{lllllll} \hline
    Authors~[Method]  & $T_{c}$ &  $U/t$  &  SC   & AF   & PS  & $\Delta E$    \\    
    \text{Present study}~[mVMC]  &   -~(GS)  &  4-12    
  &  $\delta\lesssim0.2$\footnote{{Only for $U/t \gtrsim 8$. It is absent for $U/t \lesssim 6$.} } & $\delta\lesssim0.18$  & 
     $\delta\lesssim0.19$\footnote{$U/t=10$,(SC-AFI)} & $\sim0.004t$\\
     \\ \hline
    \text{N.~Furukawa and M. Imada~(1992)~\cite{Furukawa1992}}~[QMC]  &   -~(GS)  &  4    &  No SC    & No AF& No PS & -\\
    \text{S.~Watanabe and M. Imada~(2004)~\cite{Watanabe}}~[PIRG]\footnote{$t^{\prime}/t=0$ and $-0.2$}  &   -~(GS)  &  4    &  No SC    & No AF& No PS & -\\
    \text{T.~Aimi and M. Imada~(2007)~\cite{AimiGBMC}}~[GBMC]       &   -~(GS)  &  4-6 &  No SC  & No AF& -  & -\\
    \\ \hline
    \text{T.~A.~Maier {\it et al.}(2004)\cite{Maier}}~[DCA] & $\sim0.02t$  &  4 
     &  $\delta\sim0.1$  & -& - & -      \\
    \text{E.~Khatami {\it et al.}(2010)~\cite{Khatami}}~[DCA]  &   $\sim0.02t$  &  8 &  
    $\delta\lesssim0.2$ & -&QCP at $\delta\sim0.9$  & -\\
    \text{M.~Capone and G.~Kotliar (2006)~\cite{Capone}}~[CDMFT] & -~(GS) &  4-16 
     &$\delta\lesssim$0.15\footnote{$U/t=4$-$16$}
     &$\delta\lesssim0.15$    
     &$0.05\lesssim\delta\lesssim0.15$\footnote{$U/t=16$,~(AF-SC)} & $\sim0.01t$\footnote{SC-AFM, $U/t=16$}   \\
    \text{M.~Aichhorn {\it et al.}(2007)~\cite{Aichhorn}}~[VCA]
      \footnote{$t^{\prime}/t=-0.3$, where $t^{\prime}$ is the next-nearest-neighbor transfer.}  
       &   -~(GS)   &  4-12   
       &$\delta\lesssim0.2$\footnote{$U/t=4$-$12$} 
       &$\delta\lesssim0.15$      
       &$0.05\lesssim\delta\lesssim0.15$\footnote{$U/t=8$, (SC-SC+AF)}     \\
    \text{E.~Gull  {\it et al.}(2012)~\cite{Gull}}~[DCA] &   $\sim0.016t$ &  4-6.5  
       &  $\delta\lesssim$0.15\footnote{$U/t=4$-6.5}  & -& - & $\sim 0.01t$           \\
    \text{G.~Sordi {\it et al.}(2012)~\cite{Sordi}}~[CDMFT] &$\sim0.02t$&5.2-6.2
     & $\delta\lesssim0.08$\footnote{$U/t=5.2$-6.2} & -& $0.04\lesssim\delta\lesssim0.06$\footnote{(Metal-Metal)} \\
    \\ \hline
    \text{{T.~Giamarchi and C. Lhuillier (1991)}~\cite{Giamarchi_PRB1991}}~[VMC]  & -~(GS) & 10 
     & $\delta\lesssim0.4$ &$\delta\lesssim0.2$\footnote{SC+AF} 
     & -  & - \\
    \text{H.~Yokoyama {\it et al.}(2004,2012)~\cite{Yokoyama2004,Yokoyama2012}}~[VMC]  & -~(GS) & 0-30 
     & $\delta\lesssim0.2$\footnote{$U/t=5$-30} &$\delta\lesssim0.15$ 
     & $\delta\lesssim0.1$  & $\sim0.01t$ \\
    \text{D.~Eichenberger and D. Baeriswyl~(2009)~\cite{Eichenberger}}~[VMC] &-~(GS)&6 
     &  $\delta\lesssim0.2$ &$\delta\lesssim0.1~?$    & -  & $\sim 0.01t$\\
    \\ \hline
    \text{E.~Neuscamman {\it et al.}(2012)~\cite{Neuscamman}}~[VMC]\footnote{$N_{\rm s}=8\times8$, TABC} &   -~(GS)   
    &  4    &  -  & -   & 
    $\delta\lesssim0.15$\footnote{(Metal-AFI?)}& - \\ 
    \text{S.~Zhang {\it et al.}(1997)~\cite{CPMC}}~[CPMC]\footnote{$N_{\rm s}$$\leq16\times16$, PP}  
      &   -~(GS)   &  2-8    &  No SC&  -  
      & - & -\\
    \text{C.~-C.~Chang {\it et al.}(2008,2010)~\cite{Chang,Chang2010}}~[CPMC]\footnote{$N_{\rm s}=8\times8$~-~$16\times16$, TABC}  
      &   -~(GS)   &  2-12    &  -& $\delta\lesssim0.1$\footnote{Incommensurate spin structures.}   
      & $\delta\lesssim0.1$\footnote{For $U/t\geq8$, spatially inhomogeneous state is obtained.} & -\\
    \text{S.~Sorella (2011)~\cite{Sorella}}~[VMC]\footnote{$N_{\rm s}=98$}  
      &   -~(GS)   &  4    &  -& -
      & No PS & -\\
    {\text{F.~Becca {\it et al.}~(2000)~\cite{Becca}}~[GFMC]} &   -~(GS)  
    &  4-10   &- & -  &  No PS  & -    \\ 
    \text{{L.~F.~Tocchio {\it et al.}~(2013)}~\cite{Tocchio}}~[VMC] &-~(GS)&6
     &  - & -    & No PS  & - \\
    \\ \hline
  \end{tabular}}
  \end{flushleft}
\caption{List of previous studies on the doped Hubbard model.
 SC, AF (AFM/AFI), PS, {TABC}, and GS represent 
 superconductivity, antiferromagnetic (antiferromagnetic metal/insulator), 
 phase separation, {twist-averaged boundary condition}, and ground state, respectively.}
\label{Table:Review}
\end{table*}%
 In \tm{Table}~\ref{Table:Review}, we summarize the previous numerical studies on the doped Hubbard model. 
 We summarize estimates of $T_{c}$, 
 region of superconducting (SC) phase, 
 and antiferromagnetic (AF) phase.
 We also summarize information on phase separation (PS) and
 condensation energy $\Delta E$.

In the first column, the results of the present study is summarized.

In the second column, we show {several Monte Carlo i.e.
auxiliary-field quantum Monte Carlo (QMC) and Gaussian-basis quantum Monte Carlo (GBMC)
as well as path-integral renormalization group (PIRG)} calculations.
{We note that these methods  do not restrict the form of the wavefunction {\it a priori} and give the 
accurate estimates of the energy among various numerical schemes, if the interaction is from weak to 
intermediate coupling region ($U/t\lesssim 6$). The accuracy of the PIRG has been benchmarked 
to be accurate~\cite{KashimaImada} and applied to 
various cases~\cite{KashimaImada,MoritaImada,MizusakiImada}. The GBMC has been 
benchmarked with the pre-projection method~\cite{AimiGBMC}, which substantially 
relaxes the limitation and eliminates the origin of the errors 
(boundary terms)~\cite{Corboz} and then gives good agreement with the QMC results.}
The (high-$T_{c}$) superconducting phase 
 does not appear in the region {of} $U/t \lesssim 6$ in all of these methods. {The absence is consistent with the present mVMC result, i.e.,
 we confirm that the superconducting phase {is not stabilized} for  $U/t\lesssim 6$ as shown in
 the first column. This is consistent with some other 
 results~\cite{Yokoyama2004,Eichenberger} as well as the CPMC studies\cite{CPMC}}. 

At $U/t=4$, the divergence of the compressibility is suggested at $\delta\sim 0$~\cite{Furukawa1992,Sorella}, which means that the phase separation is absent but 
the system is on the marginal quantum critical point~\cite{ImadaMQCP,MisawaImada}.  
The absence of the phase separation or restriction at most to a tiny region $\delta<0.06$
~\cite{Furukawa1992,Furukawa93,Moreo,Watanabe,Sorella} is well consistent with the present study.
The phase separation is clearly observed in a wide region of the doping concentration in the present study for the strong coupling region ($U/t> 6$), which has not been well studied before in the quantitatively accurate methods. 

 In the third column, we mainly show the results obtained 
 by dynamical mean-field theory (DMFT) calculations 
 with cluster extension such as dynamical cluster approximation (DCA) and cellular DMFT (CDMFT).  
 We also show the results of variational cluster approximations (VCA).
 
All of these works suggest that the $d$-wave superconducting phase appear
 around $\delta\sim 0.1$.
The absence of the superconductivity for $U\leq 6$ observed in the present study is not 
 consistent with DMFT and its extensions~\cite{Maier,Capone,Aichhorn,Gull,Sordi}, which may be 
 attributed to the overestimate of the superconductivity in DMFT because of 
 the mean-field approximation. We note that the presence or absence of the superconductivity is 
 determined only by the long-ranged part of the pairing correlation, while such spatial correlations and fluctuations are not captured by the DMFT.

Some works suggest that
 the first-order phase transition between {two metal phases} occurs, i.e.,
 phase separation occurs between metals~\cite{Khatami}.
 This type of phase separation is only found in DMFT calculations 
 and not observed in other calculations
 such as VMC and constrained-path Monte Carlo (CPMC) as shown in the fourth and fifth columns.

 In the fourth column, we show several previous VMC calculations.
 {In the previous VMC calculations, 
 the form of wave functions is limited and
 they use different wavefunctions
 to describe the Fermi liquid, antiferromagnetic phase, $d$-wave superconducting phase, and 
 {their coexistence phase~\cite{Giamarchi_PRB1991}}, respectively.}
 {We obtain typically 5 \% lower energy compared to early VMC results~\cite{Yokoyama2004}.
For example,  for $U/t=10$,~$L=10$,~$\delta=0.88$,~and AP boundary conditions, 
Yokoyama $et$ $al$.~\cite{Yokoyama2004} obtain $E/N_{\rm s}\sim-0.60t$ while we obtain $E/N_{\rm s}\sim-0.625t$. 
Recent VMC studies implemented a number of additional improvements to reach better accuracy,~\cite{Sorella,Tocchio} which are comparable to the present study in energy.
In contrast to most of earlier studies,
 we employ flexible one-body part of the wave functions defined in Eq.~(\ref{Eq:onebody}).
 By optimizing the long-range part of $f_{ij}$, this wave function
 can describe from insulators to antiferromagnetic metals, superconducting phases,  
 strongly correlated metals and their competitions/coexistence on an equal footing in a single framework.
 It is important for VMC results to benchmark the accuracy by comparing with the available accurate results obtained without assuming biased forms of wavefunctions as those listed in the second column.
By comparing with established results, we show in {Appendices B and G} that our wave functions allow 
precise estimations of physical properties.}

 In the fifth column, the results of VMC and CPMC methods, which 
 mainly study {the normal state properties and} instability toward PS, 
 are shown. 
 Neuscamman $et$ $al.~$\cite{Neuscamman} {have used a variational wave function with a large number of variational parameters,
 which is similar to ours}.
 {However, their estimate of the phase separation region in the
 doped Hubbard model extends to a larger doping concentration 
$\delta\sim 0.15$  even at $U/t=4$.  This contradicts other and present estimates.
The reason for the overestimate of the phase separation in Ref.\onlinecite{Neuscamman} is not clear enough for the moment. 
The CPMC studies also suggested the phase separation up to $\delta\sim 0.1$ at $U/t=4$~\cite{Chang} (or incommensurate antiferromagnetic order instead\cite{Chang2010} ).  
This has been criticized in Ref.\onlinecite{Sorella} by taking into account 
the coexisting antiferromagnetic and BCS guiding functions, which give more or less  the absence of the phase separation. 
{
Many works including numerically exact method such as QMC 
suggest that PS does not occur in the weak coupling region ($U/t\lesssim8$)
and our present work is consistent with them.}
{Although Becca $et$ $al.$~\cite{Becca}  claim that 
PS does not occur even in the strong coupling region ($U/t=10$) from the result of charge {structure factor} 
by using Green-function Monte Carlo (GFMC) method, 
the charge {structure factor} is not a proper quantity to detect the 
PS as we will show in Appendix E. }
For the case with the next-neighbor-hopping $t'=-0.4t$, the phase separation is observed at strong coupling $U/t=10$~\cite{Tocchio}.}

{{Our result  on the PS is consistent with the
most of the former studies where the PS occurs in the strong coupling region.}
{However, the relation between the PS
and the superconductivity clarified {as a key} in the present work 
has not been well studied in the literatures.}
}

 {Here, we mention about the previous studies on the extended Hubbard model.
 In the strong coupling region, effects of intersite 
 interactions such as $V/t$ and $J/t$ are
 studied by using VMC and CDMFT~\cite{Sorella_VJ,Tremblay_V}. 
 They showed that the intersite Coulomb interaction $V$ reduces
 the superconducting order parameter. 
 However, they do not study the competitions with other phases such as antiferromagnetic phase.
 Thus, it is not clear whether the superconducting phase is robust 
 against intersite Coulomb interactions. 
 By performing the {high-precision} calculations that treat the superconducting phase
 and antiferromagnetic phase or strongly correlated metal on an equal footing,
 we show that the superconducting phase becomes unstable for small $V$ ($V/U=0.1$).
 This 
 fragility of superconducting phase is not clarified in previous studies. 
 In addition, we again note that the CDMFT often overestimates 
 the stability of the superconducting phase because of its mean-field nature.

\section{ Benchmark of present mVMC method }
\label{sec:benchmark}

To show the accuracy of the present mVMC method,
we compare {our results with} 
those of {the exact diagonalization (ED),
auxiliary-field QMC, and GBMC for the Hubbard model on the square lattice},
{since they provide us in general with the best estimates of the energy as well as other physical properties}.
{A weak point of the QMC and GBMC methods are that they are applicable only in the region up to
the intermediate coupling. However, they give accurate energies and physical properties and are 
useful for the benchmark.  In fact, the QMC is a numerically exact method within the statistical error and the GBMC is well established to give very good agreement with the QMC and ED results in the range $U \leq 6$~\cite{AimiGBMC}.}

\begin{table}[b!]
\begin{center}
	\begin{tabular}{ccccc} \hline
	\text{Physical Properties}         &  mVMC($2\times2$)   & ED   \\    \hline 
  \text{$4\times4$(PP),$n=1$}  \\
	\text{Energy per site}             &  -0.8500(1)          &  -0.8513       \\
  \text{$S(\bm{q}_{\rm peak})/N_{\rm s}$} &  0.0575(2)            &   0.0569       \\ 
  \text{$\bm{q}_{\rm peak}$}              &  ($\pi$,$\pi$)     & ($\pi,\pi$)  \\ 
  {\text{$\langle\boldsymbol{S}_{i}\cdot\boldsymbol{S}_{j}\rangle$}}   &   -0.2063(14)    &  -0.2063 \\ 
   \\ \hline
  \text{$4\times4$(PP),$n=0.625$}  \\
	\text{Energy per site}             &   -1.2196(1)         &  -1.22380       \\
  \text{$S(\bm{q}_{\rm peak})/N_{\rm s}$} &    0.0130(1)         &   0.01300       \\ 
  \text{$\bm{q}_{\rm peak}$}              &    ($\pi/2$,$\pi$)     &   ($\pi/2,\pi$)  \\ 
  {\text{$\langle\boldsymbol{S}_{i}\cdot\boldsymbol{S}_{j}\rangle$}}   &   -0.0704(5)   &  -0.0683 \\ 
   \\ \hline
  \text{$4\times4$(AP),$n=1$}  \\
	\text{Energy per site}             &   -0.9081(1)          &  -0.9120       \\
  \text{$S(\bm{q}_{\rm peak})/N_{\rm s}$} &    0.0414(1)           &   0.039698      \\ 
  \text{$\bm{q}_{\rm peak}$}              &  ($\pi$,$\pi$)      &   ($\pi,\pi$)  \\ 
  {\text{$\langle\boldsymbol{S}_{i}\cdot\boldsymbol{S}_{j}\rangle$}}   &   -0.1591(8)   &  -0.1537 \\ 
   \\ \hline
  \text{$4\times4$(AP),$n=0.75$}  \\
	\text{Energy per site}             &   -1.1504(1)       &  -1.1607       \\
  \text{$S(\bm{q}_{\rm peak})/N_{\rm s}$} &    0.0179(2)      &   0.0179      \\ 
  \text{$\bm{q}_{\rm peak}$}              &    ($\pi$,0)      &   ($\pi,\pi/2$)  \\  
  {\text{$\langle\boldsymbol{S}_{i}\cdot\boldsymbol{S}_{j}\rangle$}}   &    -0.0944(7)   &  -0.0936 \\ \hline
  \end{tabular}
\end{center}
\caption{[$U{/t}=4$] Comparison of Energy, peak value of spin structure $S(\bm{q}_{\rm peak})/N_{\rm s}$, its wavenumber $\bm{q}_{\rm peak}$,
and nearest-neighbor spin correlation $\langle\boldsymbol{S}_{i}\cdot\boldsymbol{S}_{j}\rangle$ 
{between the exact diagonalization (ED) results and those of mVMC, where mVMC$(2\times 2)$ means that 
the number of the variational parameters for $f_{ij}$ is $2\times 2\times N_s$.}
{The parentheses denote the error bars in the last digit.}}
\label{Table:4by4_U4}
\end{table}%

\begin{table}[t!]
\begin{center}
	\begin{tabular}{cccc} \hline
	\text{Physical Properties}           &   mVMC($2\times2$)    & ED   \\    \hline 
  \text{$4\times4$(PP),$n=1$}  \\
	\text{Energy per site}               &   -0.43632(5)                 &  -0.43931     \\
  \text{$S(\bm{q}_{\rm peak})/N_{\rm s}$}   &    0.0860(3)                  &   0.0835       \\ 
  \text{$\bm{q}_{\rm peak}$}       &    ($\pi$,$\pi$)    &   ($\pi,\pi$)  \\ 
  {\text{$\langle\boldsymbol{S}_{i}\cdot\boldsymbol{S}_{j}\rangle$}}   &  -0.3010(9)     &  -0.3057 \\ 
   \\ \hline
  \text{$4\times4$(PP),$n=0.625$}  \\
	\text{Energy per site}                &   -1.0444(3)                & -1.0564       \\
  \text{$S(\bm{q}_{\rm peak})/N_{\rm s}$}    &    0.01505(7)              &  0.01508       \\ 
  \text{$\bm{q}_{\rm peak}$}                 &    ($\pi/2$,$\pi$)      &  ($\pi/2,\pi$)  \\ 
  {\text{$\langle\boldsymbol{S}_{i}\cdot\boldsymbol{S}_{j}\rangle$}}   & -0.0818(5)  &  -0.0754 \\ 
   \\ \hline
  \text{$4\times4$(AP),$n=1$}  \\
	\text{Energy per site}                &   -0.4422(1)               &   -0.4457       \\
  \text{$S(\bm{q}_{\rm peak})/N_{\rm s}$}    &    0.0852(2)               &    0.0819       \\ 
  \text{$\bm{q}_{\rm peak}$}                 &    ($\pi$,$\pi$)          &   ($\pi,\pi$) \\ 
  {\text{$\langle\boldsymbol{S}_{i}\cdot\boldsymbol{S}_{j}\rangle$}}   &  -0.2994(17)   &  -0.3044  \\ 
   \\ \hline
  \text{$4\times4$(AP),$n=0.75$}  \\
	\text{Energy per site}      &   -0.9022(3)                &  -0.9255       \\
  \text{$S(\bm{q}_{\rm peak})/N_{\rm s}$}    &    0.0261(3)      &   0.0216      \\ 
  \text{$\bm{q}_{\rm peak}$}       &    ($\pi$,0)                &   ($\pi,\pi/2$)  \\  
  {\text{$\langle\boldsymbol{S}_{i}\cdot\boldsymbol{S}_{j}\rangle$}}   &  -0.1087(15)     &  -0.1073 \\ \hline
  \end{tabular}
\end{center}
\caption{[$U{/t}=10$] Comparison of Energy, peak value of spin structure $S(\bm{q}_{\rm peak})/N_{\rm s}$, its wavenumber $\bm{q}_{\rm peak}$, and
 nearest-neighbor spin correlation $\langle\boldsymbol{S}_{i}\cdot\boldsymbol{S}_{j}\rangle$. {The method is the same as Table \ref{Table:4by4_U4}}.
{The parentheses denote the error bars in the last digit.}}
\label{Table:4by4_U10}
\end{table}%

\begin{table}[h!]
\begin{center}
  \begin{tabular}{ccccc} \hline
                                  &   QMC           &  GBMC        & mVMC   \\    \hline
    \text{$8\times8$~(PP), $n=50/64$}  \\
    \text{$U/t=4$}                  &   -72.80(6)        &  -72.51(5)   & -71.417(4) \\
    \text{$U/t=6$}                  &   -                &  -63.64(12)  & -62.553(9) \\
    \\ \hline
    \text{$10\times10$~(PP), $n=82/100$}  \\
    \text{$U/t=4$}                  &   -109.7(6)        &   -          & -107.51(1) \\
    \text{$U/t=6$}                  &   -                &  -92.07(22)  & -91.91(1) \\
    \\ \hline
    \text{$12\times12$~(PP), $n=122/144$}  \\
    \text{$U/t=4$}                  &   -151.4(14)        &   -          & -150.14(2) \\ \hline
  \end{tabular}
\end{center}
\caption{Comparison of total energy {between mVMC results and those of numerically well benchmarked accurate methods}.
{The parentheses denote the error bars in the last digit.}}
\label{Table:compEne}
\end{table}%

In \tm{Tables}~\ref{Table:4by4_U4} and \ref{Table:4by4_U10} , we show the results of 
mVMC and ED at half filling as well as
doped case for $U/t=4$ and $U/t=10$.
To see the boundary effects, we calculate both {PP and
 AP boundary conditions}.
For the doped case, we choose the closed-shell filling for PP and AP boundary conditions.
Total energy is well consistent with the values of ED and
its relative errors $\delta E= 1-E_{\rm mVMC}/E_{\rm ED}$ are 
typically less than 1\% even for the strong coupling regime ($U/t=10$). 
Peak values of {the} spin structure factor are also well consistent with
the exact values in all {the} cases.
{We also confirm that nearest-neighbor spin correlations 
$\langle\boldsymbol{S}_{i}\cdot\boldsymbol{S}_{j}\rangle$ are well consistent with the results of ED.}

{
We \tm{also} perform the first- and second-step power Lanczos method
for $U/t=4$ at half filling. In Fig.~\ref{fig:var_ext},
we plot the energy as a function of the variance, which is 
defined as $\Delta_{\rm var}=(\la H^2\ra-\la H\ra^2)/\la H\ra^2$.
As shown in Fig.~\ref{fig:var_ext}, 
the power Lanczos steps systematically \tm{improve} the energies.
Since the energy difference from the exact ground-state energy 
is linearly proportional to $\Delta_{\rm var}$ for
sufficiently small variance~\cite{Sorella_PRB2001},
we can estimate \tm{more precise} ground-state energy by performing the linear
fitting of the energies as a function of $\Delta_{\rm var}$.
}
{
Since the studies with the Lanczos step require substantially heavier
computational costs and the physical quantities do change little after the
Lanczos step as in \tr{{Figs.~\ref{fig:SCNe110} and \ref{fig:1stLz}}, we have performed the Lanczos calculation
only for a small number of examples, which is sufficient to confirm the validity
of the result. Systematic studies of the effects of further power Lanczos steps
are beyond the scope of this paper and left for future studies.
}

\begin{figure}[htb!]
	\begin{center}
		\includegraphics[width=8cm,clip]{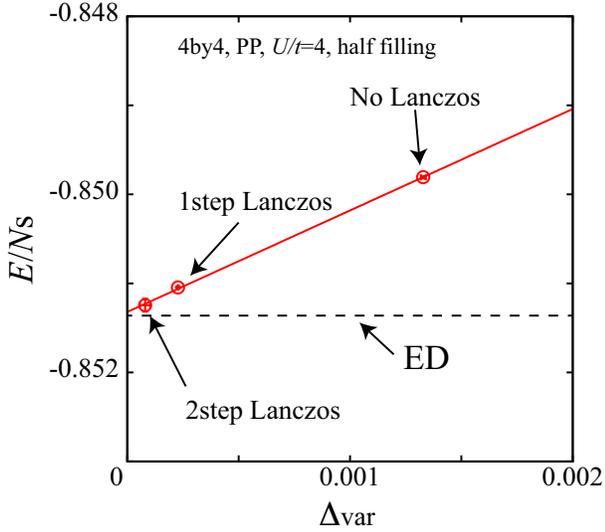}   
	\end{center}
\caption{ Variance $\Delta_{\rm var}$ dependence of energies for zero-, first-, and second-step 
power Lanczos calculations. Solid line represents the result of linear fitting of energies.
We employ PP boundary condition.}
\label{fig:var_ext}
\end{figure}%

In \tm{Table}~\ref{Table:compEne}, we compare the results of 
mVMC with available QMC and GBMC  at different fillings for $U/t=4$ and $U/t=6$.
The PP boundary condition is employed.
In large systems, 
our mVMC offers {consistent} results with the {QMC and GBMC} calculations.
{These} results also confirm the accuracy of our mVMC method.
{We also note that the accuracy of the GBMC compared with the available 
QMC results has well been benchmarked in 
physical properties including the superconducting correlations~\cite{AimiGBMC}.}

In Figs. \ref{fig:Ne10PPSC} and \ref{fig:Ne12PASC},
we show {the pairing correlations} $P_{d_{x^2-y^2}}(r)$ calculated by mVMC and ED
for doped case. Our mVMC method well reproduces the 
exact superconducting correlation for all the distances.
We note that the deviation from the exact value is large for 
$U/t=10$ at $r=\sqrt{2}$ in Fig.~\ref{fig:Ne10PPSC}.
This deviation of short-range part is not 
significant because the long-range part of
$P_{d_{x^2-y^2}}(r)$ is essential to detect the appearance of
superconducting phase.
{For larger system size ($N_{\rm s}=8\times8$), we compare the pairing correlations obtained by mVMC
with those by GBMC. As shown in Fig.~\ref{fig:8by8},
our mVMC method well reproduces the
exact superconducting correlation 
for all the distances.}

{We also show doping dependence of the spin structure factor
$S(\bm{q}_{\rm peak})$ for $U/t=4$ in Fig.~\ref{fig:Sq}.
Our mVMC well reproduces the QMC results.}
{The accuracy and applicability of the mVMC method in general have also been
examined in the literature~\cite{TaharaVMC_Full,Becca2009,misawa2012,VMC1D}.}

\begin{figure}[htb!]
	\begin{center}
		\includegraphics[width=8cm,clip]{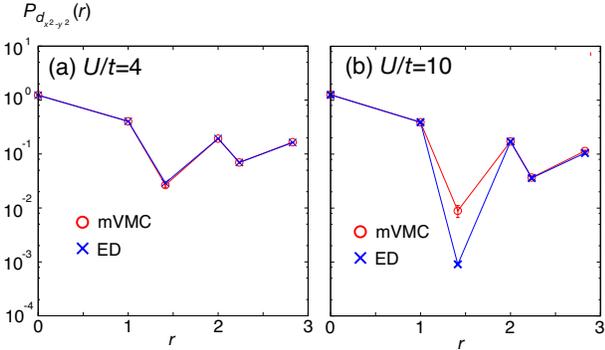}   
	\end{center}
\caption{Distance dependence of $d_{x^{2}-y^{2}}$-wave superconducting correlation $P_{d_{x^2-y^2}}(r)$ at
$n=10/16=0.625$ for PP boundary condition. For $U/t=4$ and $U/t=10$, mVMC well reproduces the exact values.
In the present plots and the plots in the later figures,
the error bars indicate the estimated statistical errors of the Monte Carlo sampling (See Sec.~II) .}
\label{fig:Ne10PPSC}
\end{figure}%

\begin{figure}[htb!]
	\begin{center}
		\includegraphics[width=8cm,clip]{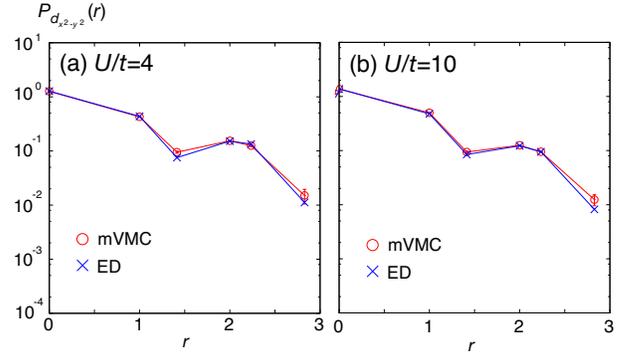}   
	\end{center}
\caption{{Superconducting correlation $P_{d_{x^2-y^2}}(r)$ for $d_{x^{2}-y^{2}}$-wave symmetry as a function of distance $r$} at
$n=12/16=0.75$ {for $4\times 4$ lattice with} AP boundary condition. For {both} $U/t=4$ and $U/t=10$, mVMC well reproduces the exact values {(ED)}.}
\label{fig:Ne12PASC}
\end{figure}%

\begin{figure}[htb!]
	\begin{center}
		\includegraphics[width=7cm,clip]{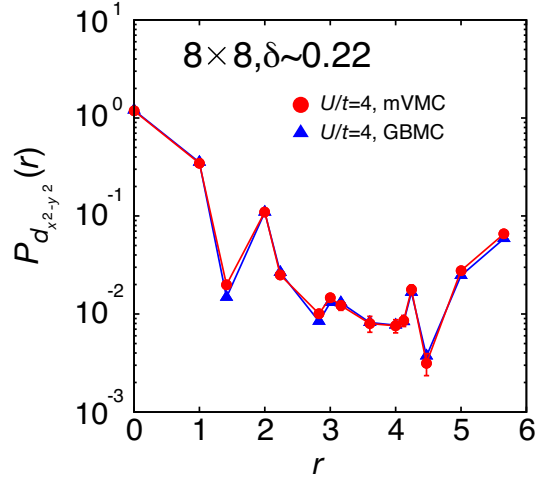}   
	\end{center}
\caption{Superconducting correlation $P_{d_{x^2-y^2}}(r)$ for $d_{x^{2}-y^{2}}$-wave 
symmetry as a function of distance $r$ for
$\delta=1-50/64\sim0.22$ and $U/t=4$ at $N_{\rm s}=8\times 8$ (PP boundary condition).
It is confirmed that mVMC well reproduces 
the essentially exact results of GBMC. }
\label{fig:8by8}
\end{figure}%

\begin{figure}[htb!]
	\begin{center}
		\includegraphics[width=7cm,clip]{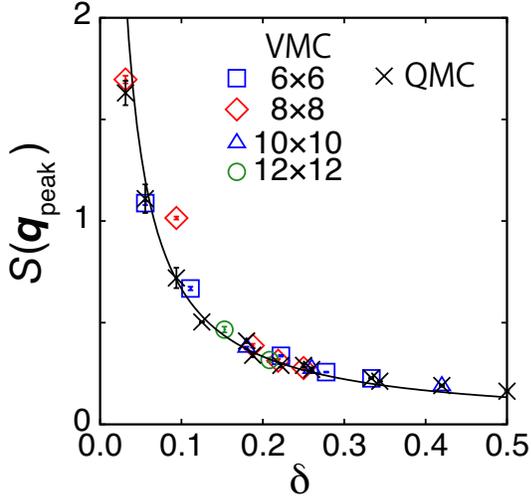}   
	\end{center}
\caption{{Doping dependence of spin structure factor $S(\bm{q}_{\rm peak})$ for $U/t=4$ and several different
system sizes (PP boundary condition).
QMC results~\cite{Furukawa1992} are shown by black crosses.
Black solid line is guide for eyes.}}
\label{fig:Sq}
\end{figure}%


\section{ Details of condensation energy}
\label{sec:condene}
In this section, we show the details of condensation energy, i.e.,
kinetic-energy gain $\Delta E_{\rm kin}$ and potential-energy gain $\Delta E_{U }$,
which are defined as
\begin{align}
 E_{\rm kin} &= {-t}\sum_{\langle i,j\rangle}\langle c_{i\sigma}^{\dagger}c_{j\sigma}+{\rm h.c.} \rangle, \\ \notag
 E_{U} &= {U}\sum_{i}\langle n_{i\uparrow}n_{i\downarrow}\rangle, \\ \notag
 \Delta E_{\rm kin}   &= {(E_{\rm kin,SC}-E_{\rm kin,Normal})}/{N_{\rm s}}, \\ \notag
 \Delta E_{U}   &= {(E_{\rm {\it U},SC}-E_{\rm {\it U},Normal})}/{N_{\rm s}}. \notag
\end{align}
We also show the nearest-neighbor spin correlation 
$\Delta S$, which is defined as
\begin{align}
  S_{\rm nn} &= \langle \boldsymbol{S}_{i}\cdot \boldsymbol{S}_{j}\rangle, \\
 \Delta S   &= (S_{\rm nn,SC}-S_{\rm nn,Normal}), \notag
\end{align}
where $i$ and $j$ represent the {nearest} neighbor sites.

\begin{figure}[b!]
	\begin{center}
		\includegraphics[width=7cm,clip]{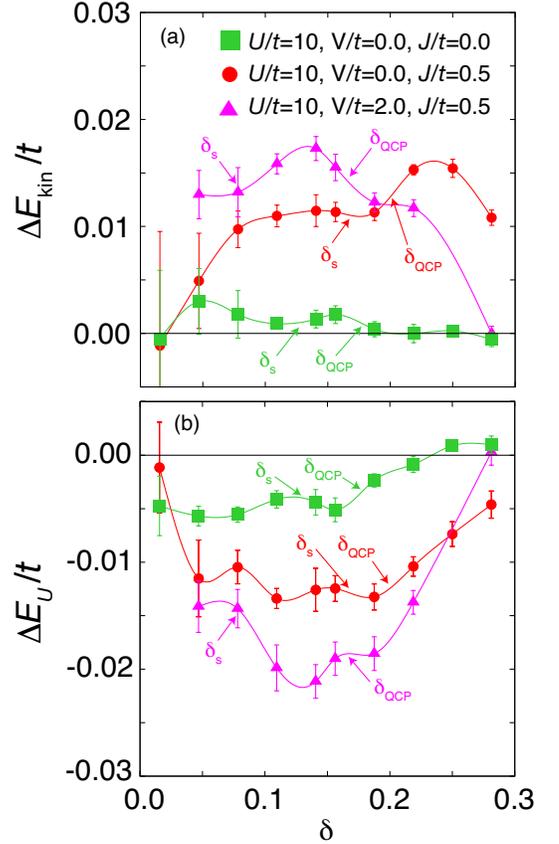}   
	\end{center}
\caption{Doping dependence of {kinetic- and potential-energy gains} in superconducting phase. 
 }
\label{fig:EkinEU}
\end{figure}%

\begin{figure}[t!]
	\begin{center}
		\includegraphics[width=7cm,clip]{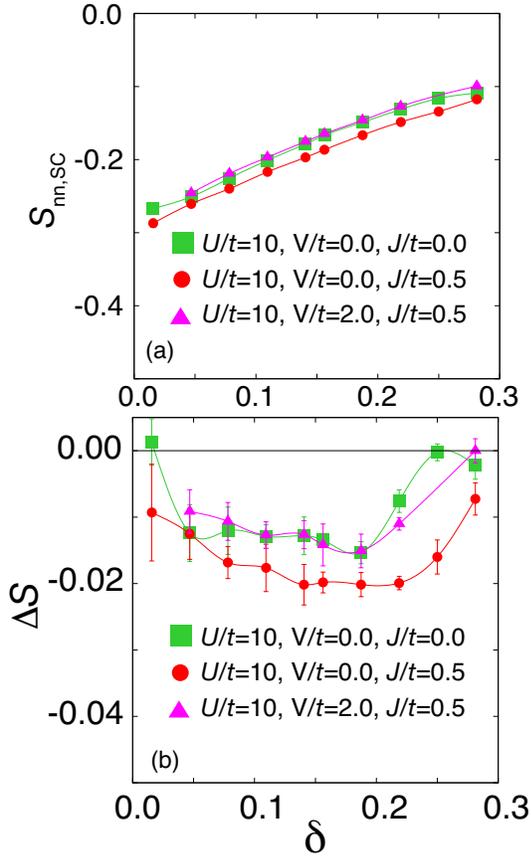}   
	\end{center}
\caption{Doping dependence of  (a) nearest-neighbor spin
correlations in superconducting phase ($S_{\rm nn,SC}$), 
and (b) $\Delta S=S_{\rm nn,SC}-S_{\rm nn,Normal}$. }
\label{fig:S}
\end{figure}%

In Fig.~\ref{fig:EkinEU}, we show doping dependence of $\Delta E_{\rm kin}$ and $\Delta E_{U}$ for
several {choices of} parameters. In the simple Hubbard model, i.e., without $V$ and $J$, 
the superconducting phase is stabilised by 
the energy gain of the potential energy {in the} whole doping region.
By introducing $V$ and $J$, the energy gain of 
potential energy becomes large while
the energy loss of kinetic energy also becomes large.
{This is because stronger pairing disturbs the single-particle 
motion and at the same time the $d$-wave pairing strictly excludes the double occupation of the 
paired electron by symmetry, which contribute to the gain in the interaction energy and the loss in the kinetic energy.  
It was claimed that the kinetic energy gain exists in the strong coupling region\cite{Yokoyama2004,Gull}. 
However, this gain was calculated in the superconducting state
without the antiferromagnetic order or correlations, while in 
reality the superconducting phase is largely coexisting with the 
antiferromagnetic order or at least with its well developed 
short-range correlations in the ground state. This coexistence leads to 
a large gain in the interaction energy and the loss in the kinetic energy 
in the superconducting state in comparison to the state with the antiferromagnetic correlations only.}   
Because {the} energy gain {arising from the} short-range 
singlet correlation exists for finite $J$,
total condensation energy becomes 
large compared to the simple Hubbard model.
As shown in Fig.~\ref{fig:S}, short-range 
singlet correlation does not largely depend on 
interaction parameters. 

{In Fig.~\ref{fig:muN_2}, we show the kinetic (potential) part of chemical potential $\mu_{\rm kin}$ ($\mu_{U}$) for $U/t=10$,
defined as 
\begin{align*}
\mu_{\rm kin}(\bar{N})&=\{E_{\rm kin}(N_{1})-E_{\rm kin}(N_{2})\}/\{N_{1}-N_{2}\}, \\
\mu_{U}(\bar{N})&=\{E_{U}(N_{1})-E_{U}(N_{2})\}/\{N_{1}-N_{2}\}{-\frac{U}{2}},
\end{align*}
where $\bar{N}=\{N_{1}+N_{2}\}/2$.
Kinetic part of chemical potential shows the convex doping dependence, 
while $\mu_{U}$ is nearly independent of the doping.
This convex doping dependence of $\mu_{\rm kin}$ suggests
that PS is mainly caused by the kinetic energy. 
}

{
A strong crossover from the 
states with the Mott proximity in the underdoped region to the overdoped region
takes place in two-fold way:
One is the charge instability represented by divergence of charge compressibility at $\delta=\delta_{\rm s}$.
The other is the magnetic instability represented by divergence of
antiferromagnetic susceptibility at $\delta=\delta_{\rm QCP}$.
This ``soft" fluctuating region provides the grounds 
for the gain in the condensation energy.}

\begin{figure}[bt!]
	\begin{center}
		\includegraphics[width=6cm,clip]{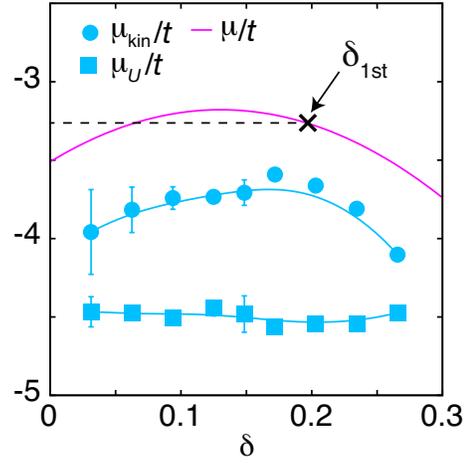}   
	\end{center}
\caption{Doping dependence of kinetic (potential) part of
chemical potential $\mu_{\rm kin}$ ($\mu_{U}$)
for $U/t=10$ and $N_{\rm s}=16\times 16$.
Solid lines are guides for eyes.
We also show total chemical potential $\mu$ for $U/t=10$, 
which is the same one as shown in Fig.~\ref{fig:muN} in the main text.
Black dashed line represents the line 
that is used for Maxwell's construction.
For comparison, we shift $\mu_{\rm kin}$ 
by {$-U/2$}. }
\label{fig:muN_2}
\end{figure}%

\begin{figure}[hbt!]
	\begin{center}
		\includegraphics[width=7cm,clip]{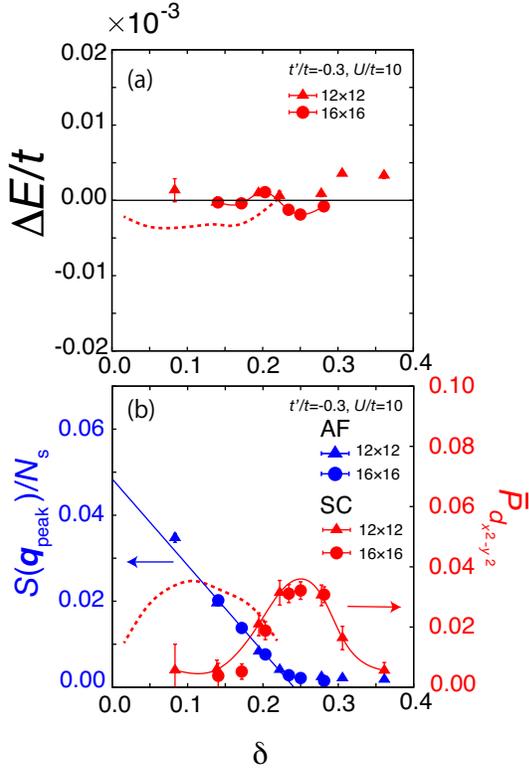}   
	\end{center}
\caption{
(a)~Doping dependence of condensation energy $\Delta E$ for $U/t=10, t^{\prime}/t=-0.3$.
Broken red line represents the condensation energy for   $U/t=10, t^{\prime}/t=0$.
(b)~Doping ($\delta$) dependence of
averaged $d_{x^{2}-y^{2}}$-wave
superconducting correlations $\bar{P}_{x^2-y^2}$
and peak values of spin structure factors  $S(\bm{q}_{\rm peak})$
{for $U/t=10$ and $t^{\prime}/t=-0.3$}.
For comparison, we plot  $\bar{P}_{x^2-y^2}$ of $U/t=10, t^{\prime}/t=0$
by broken line.
}
\label{fig:tp03}
\end{figure}%

\section{ Results with next-nearest-neighbor hopping $t'=-0.3t$}
\label{sec:tp}

{In this section we examine the effects of the next-nearest-neighbor hopping.
To directly compare with the case of $t^{\prime}/t=0$, we employ the same onsite Coulomb repulsion, i.e., $U/t=10$.}
When the next-nearest-neighbor hopping $t'=-0.3t$ is present following the 
realistic parameter of the cuprate superconductors, the condensation energy 
is strongly suppressed as we see in Fig.~\ref{fig:tp03}~(a). Concomitantly with this suppression, 
the phase separation also disappears as we see Fig.~\ref{fig:tp03muN}.  The antiferromagnetically
ordered region changes little as we see in Fig.~\ref{fig:tp03}~(b). The results are not well consistent with the 
experimental results of the hole doped copper oxides expected from the material dependence of the parameters
in the following points: (1) The suppression of the superconductivity at larger $-t'/t$ does not follow the
relation between the expected material dependence of $t'/t$ and the critical temperature $T_c$~\cite{Andersen}.  
(2) Wide  antiferromagnetically ordered region is not consistent with a quick 
destruction of the antiferromagnetic order upon hole doping.  
The origin of the discrepancy is not clear at the moment. 
Possible origins are the following: 
{(1) Realistic value of the onsite Coulomb repulsion is smaller than the present value $U/t=10$.}
(2) A combination of $V$ and $J$ expected 
in the effective low-energy model is required to stabilize the superconductivity. 
(3) Single band models are not sufficient to reproduce the quantitative aspect of the copper oxides. 
{(4) Small but finite impurities immediately destroy the antiferromagnetic order.} 

\begin{figure}[bht!]
	\begin{center}
		\includegraphics[width=6cm,clip]{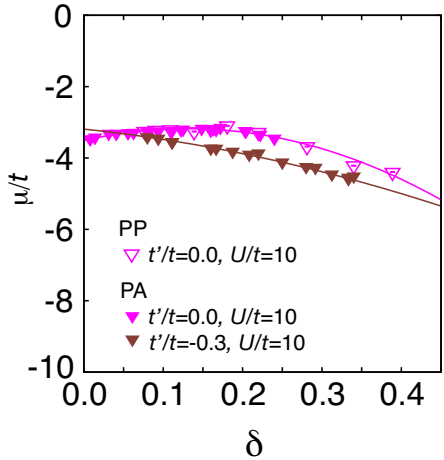}   
	\end{center}
\caption{Doping dependence of 
chemical potential $\mu_{\rm}$ 
for $t^{\prime}/t=0$ and $t^{\prime}/t=-0.3$. 
}
\label{fig:tp03muN}
\end{figure}%

\section{ Amplitude of charge structure factor in macroscopic phase-separated phase}
\label{sec:Nq}
In this section, we estimate the 
amplitude of the charge {structure factor} {allowed in finite size systems when
the phase separation occurs as a macroscopic phase.}
{In the canonical ensemble, the charge structure factor $N(\vec{q})=\frac{1}{N_{s}}\sum_{i,j}
\langle n_{i}n_{j}\rangle e^{i\bm{q}\cdot(\bm{r}_{i}-\bm{r}_{j})},$ at $\vec{q}=0$ must be zero 
{because total charge should be conserved}, 
while one may expect the growth of  $N(\vec{q})$  at the lowest possible 
wavenumber as the signature of the Bragg peak at $\vec{q}=0$ expected for the phase separation region. 
However, we here show that the growth is in practice suppressed by the 
energy loss caused by the domain wall formation in numerically accessible system sizes.}  

{Here, we first roughly estimate the energy cost caused by the density modulation imposed in a metal with the period of system size (namely at the nonzero and lowest possible wavenumber in the periodic boundary condition) 
to simulate the energy cost by the domain wall formation between two {different} 
density {phases}. (Note that this estimate is valid 
if the density modulation from the uniform phase is small, which is justified later.)} 
{For this purpose, we} consider the non-interacting Hamiltonian 
$H_{0}=\sum_{\vec{k},\sigma}\epsilon_{\vec{k}}c_{\vec{k}\sigma}^{\dagger}c_{\vec{k}\sigma}$,
where $\vec{k}$ is momentum vector and $\epsilon_{\vec{k}}$ is band dispersion, respectively.
The ground state of this Hamiltonian is Fermi-sea state {(with of course uniform density)}, which is defined as
$|\phi_{0}\ra=\prod_{|\vec{k}|<k_{F},\sigma}c^{\dagger}_{\vec{k}\sigma}|0\ra$,
where $k_{F}$ is Fermi wavenumber.
Here, we {calculate} the energy loss in the  
charge-modulated (CM) phase {$|\phi_{\rm CM}\ra$}, which is defined as 
\begin{align}
|\phi_{\rm CM}\ra &=\hat{\rho}|\phi_{0}\ra, \\
\hat{\rho}&=1+\gamma\hat{n}_{\vec{q}}, \\
\hat{n}_{\vec{q}} &=\sum_{\vec{r}_{i},\sigma}c_{\vec{r}_{i}\sigma}^{\dagger}c_{\vec{r}_{i}\sigma}e^{i\vec{q}\vec{r}_{i}} 
=\sum_{\vec{k}}c^{\dagger}_{\vec{k}+\vec{q}\sigma}c_{\vec{k}\sigma}, 
\end{align}
where 
$\vec{q}$ is the wavenumber of charge modulation.
For simplicity, we consider square lattice [{$\epsilon_{\vec{k}}=-2t^*(\cos{k_{x}+\cos{k_{y}}})$}],
$\vec{q}=(q_{x}=2\pi/L,0)$~($L$ is the linear dimension of system), and half filling (see Fig.~\ref{fig:PS}(a)).
{Note that $\vec{q}$ is the lowest possible wavenumber of the density modulation.}
First, we calculate the local density {at site $l$} as follows:
\begin{align}
\langle c_{l\sigma}^{\dagger}c_{l\sigma}\rangle &
=\frac{\la \phi_{\rm CM}|c_{l\sigma}^{\dagger}c_{l\sigma}|\phi_{\rm CM}\ra }{\la \phi_{\rm CM}|\phi_{\rm CM}\ra}\\
&=\frac{N_{e}}{2N_{s}}+\frac{2\gamma}{L}\frac{\cos{\frac{2\pi l}{L}}}{1+M|\gamma|^2},
\end{align}
where $N_{e}$ is number of total electrons and
$M=\la \phi_{0}|\hat{n}_{\vec{q}}^{\dagger}\hat{n}_{\vec{q}}|\phi_{0}\ra=\sum_{\vec{k}\in R,\sigma}\sim 2L$ 
(definition of $R$, see Fig.~\ref{fig:PS}(a)).
{Therefore, by assuming $M|\gamma|^2\ll 1$, amplitude of charge modulation $\eta$ is 
approximately given as 
\begin{align}
\eta \sim 2\times\frac{2\gamma}{L},
\end{align}
where factor $2$ comes from the spin degrees of freedom.}
Here, we define mean charge modulation $\bar{\eta}$ as 
\begin{align}
\bar{\eta} =\frac{1}{L}\times\int_{0}^{L}\eta\Big|\cos{\frac{2\pi}{L}x}\Big|dx=\frac{2}{\pi}\eta. 
\end{align}
Then, the energy {loss} within the first-order with respect to 
$\vec{q}$ is calculated as follows:
\begin{align}
E_{\vec{q}}&=\frac{\la \phi_{\rm CM}|H_{0}|\phi_{\rm CM}\ra }{\la \phi_{\rm CM}|\phi_{\rm CM}\ra}\\
&=\sum_{\vec{k}\in D,\sigma}\epsilon_{\vec{k}}
+\frac{|\gamma|^{2}}{1+M|\gamma|^{2}}\Big[-\sum_{\vec{k}\in R,\sigma}\epsilon_{\vec{k}}
+\sum_{\vec{k}\in R,\sigma}\epsilon_{\vec{k}+\vec{q}}\Big]\\
&\sim\sum_{\vec{k}\in D,\sigma}\epsilon_{\vec{k}}+
{\frac{|\gamma|^{2}q_{x}}{1+M|\gamma|^{2}}}
\sum_{\vec{k}\in R,\sigma}{\frac{\partial \epsilon_{\vec{k}}}{\partial {k_{x}}}},
\end{align}
From this, we evaluate {the energy loss arising from the density modulation},
$\Delta E_{\rm CM}$ as 
\begin{align}
\Delta E_{\rm CM} &\equiv E_{\vec{q}}-{E_{\vec{q}=\vec{0}}} \\
&=\frac{|\gamma|^{2}{q_{x}}}{1+M|\gamma|^{2}}
\sum_{\vec{k}\in R,\sigma}2t^*\sin{k_{x}} \\
&\sim16|\gamma|^2t^*, 
\end{align}
{where we again assume $M|\gamma|^{2}\ll 1$.}

\begin{figure}[hbt!]
  \begin{center}
    \includegraphics[width=7cm,clip]{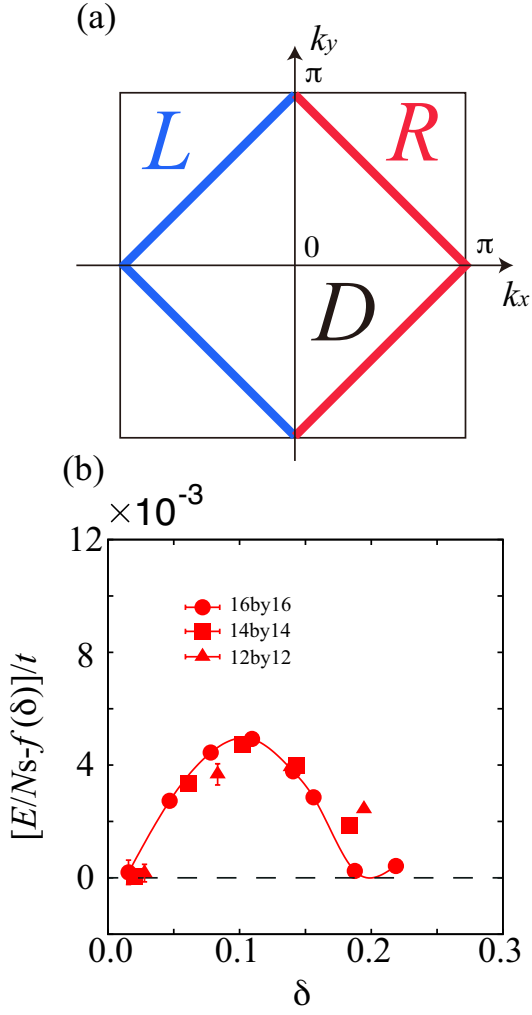}
  \end{center}
\caption{(a) Schematic picture of Fermi surface (red and blue thick line).
Region $D$ denotes the inside of the Fermi surface and 
$R (L)$ denotes the right (left) edge of the 
Fermi surface, respectively.
(b) Doping dependence of the total energy for $U/t=10, V=J=0$ for several system size. 
AP boundary condition is used. {For clarity,} we subtract $f(\delta)$, 
which is a linear function of $\delta$.
Solid line is guide for eyes.
From this, we estimate the energy gain of 
phase separation as $\Delta E_{\rm PS}=0.5\times10^{-3}t\times \delta/0.1$, 
{when the density difference of the phase separated two phases is $2\delta$.} }
\label{fig:PS}
\end{figure}%

If the energy loss $\Delta E_{\rm CM}$ is smaller than the
energy gain of the phase separation $\Delta E_{\rm PS}$,
the spatially inhomogeneous phase becomes stable.
{As shown in Fig.~\ref{fig:PS}(b), from the mVMC calculations for a typical case ($U/t=10, V=J=0$),
we evaluate the energy gain 
by the phase separation 
with the amplitude 0.1 (in the unit of the doping 
concentration $\delta$) is at most $5\times 10^{-3} t$. 
Then we have roughly estimated the 
energy gain in the case of the density modulation $\bar{\eta}$ 
as $5\times10^{-3}t\times\bar{\eta}/0.1$, simply by approximating the
curve in Fig.~\ref{fig:PS}(b) by a linear function.
}
Thus, the condition that the spatially inhomogeneous phase becomes stable  is given by  
\begin{align}
\frac{\Delta E_{\rm CM}}{N_{\rm s}}\sim \frac{16|\gamma|^2t^*}{N_{\rm s}} 
< \Delta E_{\rm PS} \sim 5\times 10^{-3}t\times\frac{\bar{\eta}}{0.1}.
\end{align}
Given this condition is satisfied {and by assuming that $t^*$ is the same as $t$}, we can evaluate the 
{maximally allowed} amplitude of the charge modulation 
as
\begin{align}
|\eta|<0.03,
\end{align}
{in finite {size} systems.}
Thus, 
{even when the phase separation is the correct solution 
in the infinite size system, the amplitude of charge {structure} 
factor $N(\vec{q})$ at the lowest possible wavenumber} for $N_{\rm s}=16\times16=256$ is given as 
\begin{align}
N({\vec{q}})&=\frac{1}{N_{\rm s}}\sum_{i,j}n_{i}n_{j}
e^{i\vec{q}(\vec{r}_i-\vec{r}_{j})}= N_{\rm s}\times |\eta|^2\sim 0.2.
\end{align}
{Although the present estimate is rough, the order estimate of 
enhancement is expected to be correct.} 
Around $\vec{q}\sim0$, we indeed see $N(\vec{q})$ in the order of 0.1
\tr{as shown in Fig.~\ref{fig:Nq}}, 
but it is buried in the background structure.
Thus, 
it is difficult to see clear signature of the phase separation
from $N(\vec{q})$  in available system size.
{In contrast to this, the doping dependence 
of the chemical potential $\mu$
offers a reliable estimation of 
phase separation region in the relatively small systems,
because they can be correctly calculated by 
the uniform density state. }
Further analysis such as performing calculations for larger system size
is intriguing issue but left for future study.


\begin{figure}[t!]
  \begin{center}
    \includegraphics[width=7cm,clip]{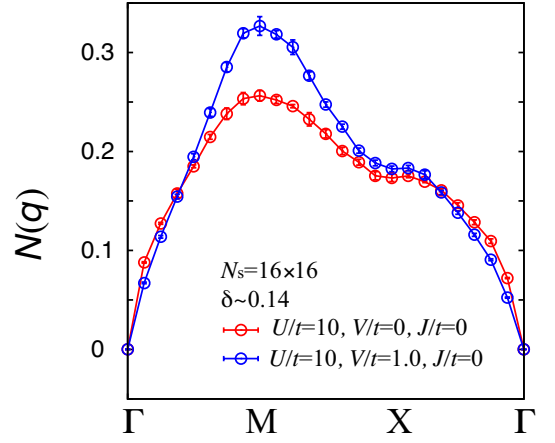}
  \end{center}
\caption{
Momentum dependence of the charge structure factor $N(\vec{q})$
at $\delta\sim0.14$ for $U/t=10, V/t=0, J/t=0$ and $U/t=10,V/t=1.0,J/t=0$.
The system size is $N_{\rm s}=16\times16$ and AP boundary conditions is employed.}
\label{fig:Nq}
\end{figure}%

\begin{figure}[t!]
  \begin{center}
    \includegraphics[width=7cm,clip]{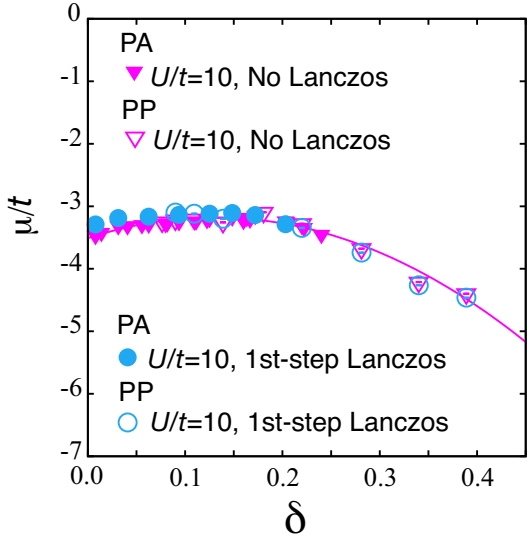}
  \end{center}
\caption{Doping dependence of chemical potential $\mu$ for $U/t=10$ after 
the first power Lanczos step. 
For comparison,
we show the doping dependence of $\mu$ for no Lanczos step.}
\label{fig:1stLz}
\end{figure}%

\section{ Doping dependence of chemical potential after Lanczos step}
Here, we show how the Lanczos step affects the doping dependence of the
chemical potential $\mu$.
In Fig.~\ref{fig:1stLz}, we show the 
doping dependence $\mu$
after the first Lanczos step for $U/t=10$ and $J=V=0$. 
From this, although Lanczos step largely improves the energies, 
we find that Lanczos step changes doping dependence of 
$\mu$ little, which is defined by the difference of the energies 
for different $\delta$ (see Eq.~\ref{Eq:mu}).
At this stage, due to the heavy numerical cost,
we can not perform the further Lanczos calculation and
systematic variance extrapolation.
Thus, precise estimation of the phase separation region 
by systematic
power Lanczos calculation is left for future study.

\section{Benchmark Results for $t$-$J$ model}
\label{sec:tJ}
The $t$-$J$ model on the  square lattice is 
defined as 
\begin{equation*}
H=-t\sum_{\la i,j\ra,\sigma}(c_{i\sigma}^{\dagger}c_{j\sigma}+{\rm h.c.})
+J\sum_{\la i,j\ra}\Big(\vec{S}_{i}\cdot\vec{S}_{j}-\frac{1}{4}n_{i}n_{j}\Big),
\end{equation*}
where the double occupancy is completely prohibited. 
In the $t$-$J$ model, 
it is suggested that the phase separation 
does not occur for sufficiently small $J$~\cite{HuBeccaSorella}.
To benchmark the accuracy of 
our variational wave function,
we perform the calculations for the $t$-$J$ model at $J/t=0.4$.
We use basically the same wave 
function defined in Eq.~(\ref{Eq:WF}) except that
we 
completely prohibit the double occupancy by using the Gutzwiller factors.
We note that the doublon-holon correlation 
factors are omitted.
We plot $e(\delta)=[E(\delta)/N_{s}-E(0)/N_{s}]/\delta$ in Fig.~\ref{fig:tJ}, which 
can be directly compared with Fig. 1 in Ref.~\onlinecite{HuBeccaSorella}.
Although values of $e(\delta)$ themselves are slightly different from 
those \tm{in Ref.~\onlinecite{HuBeccaSorella}}, 
our calculation \tm{supports the absence of} the PS \tm{consistently with Ref.~\onlinecite{HuBeccaSorella}}.
\tm{In} the Heisenberg limit ($\delta=0$),
we compare our result with the quantum Monte Carlo method~\cite{Sandvik}
and we obtain $|1-E_{\rm mVMC}/E_{\rm QMC}|\sim 0.002$ for
$N_{\rm s}=12\times12$.
This result again confirms that our 
variational wave function has
\tm{sufficient} accuracy to discuss the existence of PS.

\begin{figure}[tb!]
  \begin{center}
    \includegraphics[width=7cm,clip]{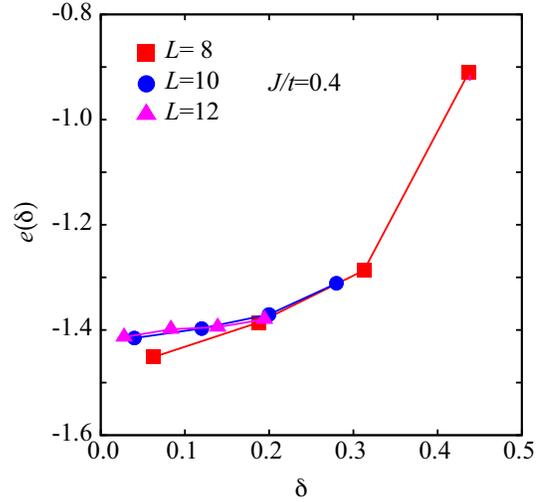}
  \end{center}
\caption{ Doping dependence of $e(\delta)$ for $J/t=0.4$.
We employ antiperiodic-periodic boundary conditions.
\tm{Since $e(\delta)$ increases monotonically, the absence of PS is concluded}~\cite{HuBeccaSorella}.}
\label{fig:tJ}
\end{figure}%

\clearpage

\begin{thebibliography}{63}
\expandafter\ifx\csname natexlab\endcsname\relax\def\natexlab#1{#1}\fi
\expandafter\ifx\csname bibnamefont\endcsname\relax
  \def\bibnamefont#1{#1}\fi
\expandafter\ifx\csname bibfnamefont\endcsname\relax
  \def\bibfnamefont#1{#1}\fi
\expandafter\ifx\csname citenamefont\endcsname\relax
  \def\citenamefont#1{#1}\fi
\expandafter\ifx\csname url\endcsname\relax
  \def\url#1{\texttt{#1}}\fi
\expandafter\ifx\csname urlprefix\endcsname\relax\def\urlprefix{URL }\fi
\providecommand{\bibinfo}[2]{#2}
\providecommand{\eprint}[2][]{\url{#2}}

\bibitem[{\citenamefont{Bednorz and M{\"u}ller}(1986)}]{Bednorz}
\bibinfo{author}{\bibfnamefont{J.~G.} \bibnamefont{Bednorz}} \bibnamefont{and}
  \bibinfo{author}{\bibfnamefont{K.~A.} \bibnamefont{M{\"u}ller}},
  \bibinfo{journal}{Z. Phys.} \textbf{\bibinfo{volume}{64}},
  \bibinfo{pages}{189} (\bibinfo{year}{1986}).

\bibitem[{\citenamefont{Mukuda et~al.}(2008)\citenamefont{Mukuda, Yamaguchi,
  Shimizu, Kitaoka, Shirage, and Iyo}}]{Mukuda}
\bibinfo{author}{\bibfnamefont{H.}~\bibnamefont{Mukuda}},
  \bibinfo{author}{\bibfnamefont{Y.}~\bibnamefont{Yamaguchi}},
  \bibinfo{author}{\bibfnamefont{S.}~\bibnamefont{Shimizu}},
  \bibinfo{author}{\bibfnamefont{Y.}~\bibnamefont{Kitaoka}},
  \bibinfo{author}{\bibfnamefont{P.}~\bibnamefont{Shirage}}, \bibnamefont{and}
  \bibinfo{author}{\bibfnamefont{A.}~\bibnamefont{Iyo}}, \bibinfo{journal}{J.
  Phys. Soc. Jpn.} \textbf{\bibinfo{volume}{77}}, \bibinfo{pages}{124706}
  (\bibinfo{year}{2008}).

\bibitem[{\citenamefont{Wu et~al.}(2013)\citenamefont{Wu, Pelleg, Logvenov,
  Bollinger, Sun, Boebinger, Vaneti\'{c}, Radovi\'{c}, and
  Bo\v{z}ovi\'{c}}}]{WuBozovic}
\bibinfo{author}{\bibfnamefont{J.}~\bibnamefont{Wu}},
  \bibinfo{author}{\bibfnamefont{O.}~\bibnamefont{Pelleg}},
  \bibinfo{author}{\bibfnamefont{G.}~\bibnamefont{Logvenov}},
  \bibinfo{author}{\bibfnamefont{A.~T.} \bibnamefont{Bollinger}},
  \bibinfo{author}{\bibfnamefont{Y.-J.} \bibnamefont{Sun}},
  \bibinfo{author}{\bibfnamefont{G.~S.} \bibnamefont{Boebinger}},
  \bibinfo{author}{\bibfnamefont{M.}~\bibnamefont{Vaneti\'{c}}},
  \bibinfo{author}{\bibfnamefont{Z.}~\bibnamefont{Radovi\'{c}}},
  \bibnamefont{and}
  \bibinfo{author}{\bibfnamefont{I.}~\bibnamefont{Bo\v{z}ovi\'{c}}},
  \bibinfo{journal}{Nat. Mat.} \textbf{\bibinfo{volume}{12}},
  \bibinfo{pages}{877} (\bibinfo{year}{2013}).

\bibitem[{\citenamefont{Furukawa and Imada}(1992)}]{Furukawa1992}
\bibinfo{author}{\bibfnamefont{N.}~\bibnamefont{Furukawa}} \bibnamefont{and}
  \bibinfo{author}{\bibfnamefont{M.}~\bibnamefont{Imada}}, \bibinfo{journal}{J.
  Phys. Soc. Jpn.} \textbf{\bibinfo{volume}{61}}, \bibinfo{pages}{3331}
  (\bibinfo{year}{1992}).

\bibitem[{\citenamefont{Aimi and Imada}(2007)}]{AimiGBMC}
\bibinfo{author}{\bibfnamefont{T.}~\bibnamefont{Aimi}} \bibnamefont{and}
  \bibinfo{author}{\bibfnamefont{M.}~\bibnamefont{Imada}}, \bibinfo{journal}{J.
  Phys. Soc. Jpn.} \textbf{\bibinfo{volume}{76}}, \bibinfo{pages}{113708} (\bibinfo{year}{2007}).

\bibitem[{\citenamefont{Yokoyama et~al.}(2013)\citenamefont{Yokoyama, Ogata,
  Tanaka, Kobayashi, and Tsuchiura}}]{Yokoyama2012}
\bibinfo{author}{\bibfnamefont{H.}~\bibnamefont{Yokoyama}},
  \bibinfo{author}{\bibfnamefont{M.}~\bibnamefont{Ogata}},
  \bibinfo{author}{\bibfnamefont{Y.}~\bibnamefont{Tanaka}},
  \bibinfo{author}{\bibfnamefont{K.}~\bibnamefont{Kobayashi}},
  \bibnamefont{and}
  \bibinfo{author}{\bibfnamefont{H.}~\bibnamefont{Tsuchiura}},
  \bibinfo{journal}{J. Phys. Soc. Jpn.} \textbf{\bibinfo{volume}{82}},
  \bibinfo{pages}{014707} (\bibinfo{year}{2013}).

\bibitem[{\citenamefont{Maier et~al.}(2005)\citenamefont{Maier, Jarrell,
  Schulthess, Kent, and White}}]{Maier}
\bibinfo{author}{\bibfnamefont{T.~A.} \bibnamefont{Maier}},
  \bibinfo{author}{\bibfnamefont{M.}~\bibnamefont{Jarrell}},
  \bibinfo{author}{\bibfnamefont{T.~C.} \bibnamefont{Schulthess}},
  \bibinfo{author}{\bibfnamefont{P.~R.~C.} \bibnamefont{Kent}},
  \bibnamefont{and} \bibinfo{author}{\bibfnamefont{J.~B.} \bibnamefont{White}},
  \bibinfo{journal}{Phys. Rev. Lett.} \textbf{\bibinfo{volume}{95}},
  \bibinfo{pages}{237001} (\bibinfo{year}{2005}).

\bibitem[{\citenamefont{Aichhorn et~al.}(2007)\citenamefont{Aichhorn, Arrigoni,
  Potthoff, and Hanke}}]{Aichhorn}
\bibinfo{author}{\bibfnamefont{M.}~\bibnamefont{Aichhorn}},
  \bibinfo{author}{\bibfnamefont{E.}~\bibnamefont{Arrigoni}},
  \bibinfo{author}{\bibfnamefont{M.}~\bibnamefont{Potthoff}}, \bibnamefont{and}
  \bibinfo{author}{\bibfnamefont{W.}~\bibnamefont{Hanke}},
  \bibinfo{journal}{Phys. Rev. B} \textbf{\bibinfo{volume}{76}},
  \bibinfo{pages}{224509} (\bibinfo{year}{2007}).

\bibitem[{\citenamefont{Khatami et~al.}(2010)\citenamefont{Khatami, Mikelsons,
  Galanakis, Macridin, Moreno, Scalettar, and Jarrell}}]{Khatami}
\bibinfo{author}{\bibfnamefont{E.}~\bibnamefont{Khatami}},
  \bibinfo{author}{\bibfnamefont{K.}~\bibnamefont{Mikelsons}},
  \bibinfo{author}{\bibfnamefont{D.}~\bibnamefont{Galanakis}},
  \bibinfo{author}{\bibfnamefont{A.}~\bibnamefont{Macridin}},
  \bibinfo{author}{\bibfnamefont{J.}~\bibnamefont{Moreno}},
  \bibinfo{author}{\bibfnamefont{R.~T.} \bibnamefont{Scalettar}},
  \bibnamefont{and} \bibinfo{author}{\bibfnamefont{M.}~\bibnamefont{Jarrell}},
  \bibinfo{journal}{Phys. Rev. B} \textbf{\bibinfo{volume}{81}},
  \bibinfo{pages}{201101(R)} (\bibinfo{year}{2010}).

\bibitem[{\citenamefont{Capone and Kotliar}(2006)}]{Capone}
\bibinfo{author}{\bibfnamefont{M.}~\bibnamefont{Capone}} \bibnamefont{and}
  \bibinfo{author}{\bibfnamefont{G.}~\bibnamefont{Kotliar}},
  \bibinfo{journal}{Phys. Rev. B} \textbf{\bibinfo{volume}{74}},
  \bibinfo{pages}{054513} (\bibinfo{year}{2006}).

\bibitem[{\citenamefont{Sordi et~al.}(2012)\citenamefont{Sordi, S\'emon, Haule,
  and Tremblay}}]{Sordi}
\bibinfo{author}{\bibfnamefont{G.}~\bibnamefont{Sordi}},
  \bibinfo{author}{\bibfnamefont{P.}~\bibnamefont{S\'emon}},
  \bibinfo{author}{\bibfnamefont{K.}~\bibnamefont{Haule}}, \bibnamefont{and}
  \bibinfo{author}{\bibfnamefont{A.-M.~S.} \bibnamefont{Tremblay}},
  \bibinfo{journal}{Phys. Rev. Lett.} \textbf{\bibinfo{volume}{108}},
  \bibinfo{pages}{216401} (\bibinfo{year}{2012}).

\bibitem[{\citenamefont{Gull and Millis}(2012)}]{Gull}
\bibinfo{author}{\bibfnamefont{E.}~\bibnamefont{Gull}} \bibnamefont{and}
  \bibinfo{author}{\bibfnamefont{A.~J.} \bibnamefont{Millis}},
  \bibinfo{journal}{Phys. Rev. B} \textbf{\bibinfo{volume}{86}},
  \bibinfo{pages}{241106} (\bibinfo{year}{2012}).

\bibitem[{\citenamefont{Chen et~al.}(2013)\citenamefont{Chen, Meng, Yang,
  Pruschke, Moreno, and Jarrell}}]{Chen_2013}
\bibinfo{author}{\bibfnamefont{K.-S.} \bibnamefont{Chen}},
  \bibinfo{author}{\bibfnamefont{Z.~Y.} \bibnamefont{Meng}},
  \bibinfo{author}{\bibfnamefont{S.-X.} \bibnamefont{Yang}},
  \bibinfo{author}{\bibfnamefont{T.}~\bibnamefont{Pruschke}},
  \bibinfo{author}{\bibfnamefont{J.}~\bibnamefont{Moreno}}, \bibnamefont{and}
  \bibinfo{author}{\bibfnamefont{M.}~\bibnamefont{Jarrell}},
  \bibinfo{journal}{Phys. Rev. B} \textbf{\bibinfo{volume}{88}},
  \bibinfo{pages}{245110} (\bibinfo{year}{2013}).

\bibitem[{\citenamefont{Moriya and Ueda}(2003)}]{MoriyaUeda}
\bibinfo{author}{\bibfnamefont{T.}~\bibnamefont{Moriya}} \bibnamefont{and}
  \bibinfo{author}{\bibfnamefont{K.}~\bibnamefont{Ueda}},
  \bibinfo{journal}{Rep. Prog. Phys.} \textbf{\bibinfo{volume}{66}},
  \bibinfo{pages}{1299} (\bibinfo{year}{2003}).

\bibitem[{\citenamefont{Eichenberger and Baeriswyl}(2009)}]{Eichenberger}
\bibinfo{author}{\bibfnamefont{D.}~\bibnamefont{Eichenberger}}
  \bibnamefont{and}
  \bibinfo{author}{\bibfnamefont{D.}~\bibnamefont{Baeriswyl}},
  \bibinfo{journal}{Phys. Rev. B} \textbf{\bibinfo{volume}{79}},
  \bibinfo{pages}{100510} (\bibinfo{year}{2009}).

\bibitem[{\citenamefont{Zhang et~al.}(1997)\citenamefont{Zhang, Carlson, and
  Gubernatis}}]{CPMC}
\bibinfo{author}{\bibfnamefont{S.}~\bibnamefont{Zhang}},
  \bibinfo{author}{\bibfnamefont{J.}~\bibnamefont{Carlson}}, \bibnamefont{and}
  \bibinfo{author}{\bibfnamefont{J.~E.} \bibnamefont{Gubernatis}},
  \bibinfo{journal}{Phys. Rev. Lett.} \textbf{\bibinfo{volume}{78}},
  \bibinfo{pages}{4486} (\bibinfo{year}{1997}).

\bibitem[{\citenamefont{Metzner et~al.}(2012)\citenamefont{Metzner, Salmhofer,
  Honerkamp, Meden, and Sch\"onhammer}}]{FRGRMP}
\bibinfo{author}{\bibfnamefont{W.}~\bibnamefont{Metzner}},
  \bibinfo{author}{\bibfnamefont{M.}~\bibnamefont{Salmhofer}},
  \bibinfo{author}{\bibfnamefont{C.}~\bibnamefont{Honerkamp}},
  \bibinfo{author}{\bibfnamefont{V.}~\bibnamefont{Meden}}, \bibnamefont{and}
  \bibinfo{author}{\bibfnamefont{K.}~\bibnamefont{Sch\"onhammer}},
  \bibinfo{journal}{Rev. Mod. Phys.} \textbf{\bibinfo{volume}{84}},
  \bibinfo{pages}{299} (\bibinfo{year}{2012}).

\bibitem[{\citenamefont{Lee et~al.}(2006)\citenamefont{Lee, Nagaosa, and
  Wen}}]{LeeRMP}
\bibinfo{author}{\bibfnamefont{P.~A.} \bibnamefont{Lee}},
  \bibinfo{author}{\bibfnamefont{N.}~\bibnamefont{Nagaosa}}, \bibnamefont{and}
  \bibinfo{author}{\bibfnamefont{X.-G.} \bibnamefont{Wen}},
  \bibinfo{journal}{Rev. Mod. Phys.} \textbf{\bibinfo{volume}{78}},
  \bibinfo{pages}{17} (\bibinfo{year}{2006}).

\bibitem[{\citenamefont{Emery et~al.}(1990)\citenamefont{Emery, Kivelson, and
  Lin}}]{EmeryKivelson}
\bibinfo{author}{\bibfnamefont{V.~J.} \bibnamefont{Emery}},
  \bibinfo{author}{\bibfnamefont{S.~A.} \bibnamefont{Kivelson}},
  \bibnamefont{and} \bibinfo{author}{\bibfnamefont{H.~Q.} \bibnamefont{Lin}},
  \bibinfo{journal}{Phys. Rev. Lett.} \textbf{\bibinfo{volume}{64}},
  \bibinfo{pages}{475} (\bibinfo{year}{1990}).

\bibitem[{\citenamefont{Kivelson et~al.}(1998)\citenamefont{Kivelson, Fradkin,
  and Emery}}]{Kivelson}
\bibinfo{author}{\bibfnamefont{S.~A.} \bibnamefont{Kivelson}},
  \bibinfo{author}{\bibfnamefont{E.}~\bibnamefont{Fradkin}}, \bibnamefont{and}
  \bibinfo{author}{\bibfnamefont{V.~J.} \bibnamefont{Emery}},
  \bibinfo{journal}{Nature} \textbf{\bibinfo{volume}{393}},
  \bibinfo{pages}{550} (\bibinfo{year}{1998}).

\bibitem[{\citenamefont{Chang and Zhang}(2008)}]{Chang}
\bibinfo{author}{\bibfnamefont{C.-C.} \bibnamefont{Chang}} \bibnamefont{and}
  \bibinfo{author}{\bibfnamefont{S.}~\bibnamefont{Zhang}},
  \bibinfo{journal}{Phys. Rev. B} \textbf{\bibinfo{volume}{78}},
  \bibinfo{pages}{165101} (\bibinfo{year}{2008}).

\bibitem[{\citenamefont{Sorella}(2011)}]{Sorella}
\bibinfo{author}{\bibfnamefont{S.}~\bibnamefont{Sorella}},
  \bibinfo{journal}{Phys. Rev. B} \textbf{\bibinfo{volume}{84}},
  \bibinfo{pages}{241110} (\bibinfo{year}{2011}).

\bibitem[{\citenamefont{Neuscamman et~al.}(2012)\citenamefont{Neuscamman,
  Umrigar, and Chan}}]{Neuscamman}
\bibinfo{author}{\bibfnamefont{E.}~\bibnamefont{Neuscamman}},
  \bibinfo{author}{\bibfnamefont{C.~J.} \bibnamefont{Umrigar}},
  \bibnamefont{and} \bibinfo{author}{\bibfnamefont{G.~K.-L.}
  \bibnamefont{Chan}}, \bibinfo{journal}{Phys. Rev. B}
  \textbf{\bibinfo{volume}{85}}, \bibinfo{pages}{045103}
  (\bibinfo{year}{2012}).

\bibitem[{\citenamefont{White and Scalapino}(1998)}]{Stripe_tJ}
\bibinfo{author}{\bibfnamefont{S.~R.} \bibnamefont{White}} \bibnamefont{and}
  \bibinfo{author}{\bibfnamefont{D.~J.} \bibnamefont{Scalapino}},
  \bibinfo{journal}{Phys. Rev. Lett.} \textbf{\bibinfo{volume}{80}},
  \bibinfo{pages}{1272} (\bibinfo{year}{1998}).

\bibitem[{\citenamefont{Tocchio et~al.}(2013)\citenamefont{Tocchio, Lee,
  Jeschke, Valent\'\i, and Gros}}]{Tocchio}
\bibinfo{author}{\bibfnamefont{L.~F.} \bibnamefont{Tocchio}},
  \bibinfo{author}{\bibfnamefont{H.}~\bibnamefont{Lee}},
  \bibinfo{author}{\bibfnamefont{H.~O.} \bibnamefont{Jeschke}},
  \bibinfo{author}{\bibfnamefont{R.}~\bibnamefont{Valent\'\i}},
  \bibnamefont{and} \bibinfo{author}{\bibfnamefont{C.}~\bibnamefont{Gros}},
  \bibinfo{journal}{Phys. Rev. B} \textbf{\bibinfo{volume}{87}},
  \bibinfo{pages}{045111} (\bibinfo{year}{2013}).

\bibitem[{\citenamefont{Imada}(2005{\natexlab{a}})}]{ImadaMQCP}
\bibinfo{author}{\bibfnamefont{M.}~\bibnamefont{Imada}},
  \bibinfo{journal}{Phys. Rev. B} \textbf{\bibinfo{volume}{72}},
  \bibinfo{pages}{075113} (\bibinfo{year}{2005}{\natexlab{a}}).

\bibitem[{\citenamefont{Imada}(2005{\natexlab{b}})}]{ImadaSuper}
\bibinfo{author}{\bibfnamefont{M.}~\bibnamefont{Imada}}, \bibinfo{journal}{J.
  Phys. Soc. Jpn.} \textbf{\bibinfo{volume}{74}}, \bibinfo{pages}{859}
  (\bibinfo{year}{2005}{\natexlab{b}}).

\bibitem[{\citenamefont{Gutzwiller}(1963)}]{Gutzwiller}
\bibinfo{author}{\bibfnamefont{M.~C.} \bibnamefont{Gutzwiller}},
  \bibinfo{journal}{Phys. Rev. Lett.} \textbf{\bibinfo{volume}{10}},
  \bibinfo{pages}{159} (\bibinfo{year}{1963}).

\bibitem[{\citenamefont{Jastrow}(1955)}]{Jastrow}
\bibinfo{author}{\bibfnamefont{R.}~\bibnamefont{Jastrow}},
  \bibinfo{journal}{Phys. Rev.} \textbf{\bibinfo{volume}{98}},
  \bibinfo{pages}{1479} (\bibinfo{year}{1955}).

\bibitem[{\citenamefont{Capello et~al.}(2005)\citenamefont{Capello, Becca,
  Fabrizio, Sorella, and Tosatti}}]{CapelloJastrow}
\bibinfo{author}{\bibfnamefont{M.}~\bibnamefont{Capello}},
  \bibinfo{author}{\bibfnamefont{F.}~\bibnamefont{Becca}},
  \bibinfo{author}{\bibfnamefont{M.}~\bibnamefont{Fabrizio}},
  \bibinfo{author}{\bibfnamefont{S.}~\bibnamefont{Sorella}}, \bibnamefont{and}
  \bibinfo{author}{\bibfnamefont{E.}~\bibnamefont{Tosatti}},
  \bibinfo{journal}{Phys. Rev. Lett.} \textbf{\bibinfo{volume}{94}},
  \bibinfo{pages}{026406} (\bibinfo{year}{2005}).

\bibitem[{\citenamefont{Yokoyama and Shiba}(1990)}]{YokoyamaDH}
\bibinfo{author}{\bibfnamefont{H.}~\bibnamefont{Yokoyama}} \bibnamefont{and}
  \bibinfo{author}{\bibfnamefont{H.}~\bibnamefont{Shiba}}, \bibinfo{journal}{J.
  Phys. Soc. Jpn.} \textbf{\bibinfo{volume}{59}}, \bibinfo{pages}{3669}
  (\bibinfo{year}{1990}).

\bibitem[{\citenamefont{Tahara and Imada}(2008)}]{TaharaVMC_Full}
\bibinfo{author}{\bibfnamefont{D.}~\bibnamefont{Tahara}} \bibnamefont{and}
  \bibinfo{author}{\bibfnamefont{M.}~\bibnamefont{Imada}}, \bibinfo{journal}{J.
  Phys. Soc. Jpn.} \textbf{\bibinfo{volume}{77}}, \bibinfo{pages}{114701}
  (\bibinfo{year}{2008}).

\bibitem[{\citenamefont{Gros}(1989)}]{gros1989physics}
\bibinfo{author}{\bibfnamefont{C.}~\bibnamefont{Gros}}, \bibinfo{journal}{Ann.
  Phys.} \textbf{\bibinfo{volume}{189}}, \bibinfo{pages}{53}
  (\bibinfo{year}{1989}).

\bibitem[{\citenamefont{Bajdich et~al.}(2008)\citenamefont{Bajdich, Mitas,
  Wagner, and Schmidt}}]{BajdichPRB}
\bibinfo{author}{\bibfnamefont{M.}~\bibnamefont{Bajdich}},
  \bibinfo{author}{\bibfnamefont{L.}~\bibnamefont{Mitas}},
  \bibinfo{author}{\bibfnamefont{L.~K.} \bibnamefont{Wagner}},
  \bibnamefont{and} \bibinfo{author}{\bibfnamefont{K.~E.}
  \bibnamefont{Schmidt}}, \bibinfo{journal}{Phys. Rev. B}
  \textbf{\bibinfo{volume}{77}}, \bibinfo{pages}{115112}
  (\bibinfo{year}{2008}).

\bibitem[{\citenamefont{Sorella}(2001)}]{Sorella_PRB2001}
\bibinfo{author}{\bibfnamefont{S.}~\bibnamefont{Sorella}},
  \bibinfo{journal}{Phys. Rev. B} \textbf{\bibinfo{volume}{64}},
  \bibinfo{pages}{024512} (\bibinfo{year}{2001}).

\bibitem[{\citenamefont{Heeb and Rice}(1993)}]{PowerLanczos}
\bibinfo{author}{\bibfnamefont{E.}~\bibnamefont{Heeb}} \bibnamefont{and}
  \bibinfo{author}{\bibfnamefont{T.}~\bibnamefont{Rice}}, \bibinfo{journal}{Z.
  Phys. B} \textbf{\bibinfo{volume}{90}}, \bibinfo{pages}{73}
  (\bibinfo{year}{1993}).

\bibitem[{\citenamefont{Furukwa and Imada}(1993)}]{Furukawa93}
\bibinfo{author}{\bibfnamefont{N.}~\bibnamefont{Furukwa}} \bibnamefont{and}
  \bibinfo{author}{\bibfnamefont{M.}~\bibnamefont{Imada}}, \bibinfo{journal}{J.
  Phys. Soc. Jpn.} \textbf{\bibinfo{volume}{62}}, \bibinfo{pages}{2557}
  (\bibinfo{year}{1993}).

\bibitem[{\citenamefont{Kohn and Luttinger}(1965)}]{KohnLuttinger}
\bibinfo{author}{\bibfnamefont{W.}~\bibnamefont{Kohn}} \bibnamefont{and}
  \bibinfo{author}{\bibfnamefont{J.~M.} \bibnamefont{Luttinger}},
  \bibinfo{journal}{Phys. Rev. Lett.} \textbf{\bibinfo{volume}{15}},
  \bibinfo{pages}{524} (\bibinfo{year}{1965}).

\bibitem[{\citenamefont{Chen et~al.}(1990)\citenamefont{Chen, Joynt, Zhang, and
  Gros}}]{tJAFSC_PRB1990}
\bibinfo{author}{\bibfnamefont{G.~J.} \bibnamefont{Chen}},
  \bibinfo{author}{\bibfnamefont{R.}~\bibnamefont{Joynt}},
  \bibinfo{author}{\bibfnamefont{F.~C.} \bibnamefont{Zhang}}, \bibnamefont{and}
  \bibinfo{author}{\bibfnamefont{C.}~\bibnamefont{Gros}},
  \bibinfo{journal}{Phys. Rev. B} \textbf{\bibinfo{volume}{42}},
  \bibinfo{pages}{2662} (\bibinfo{year}{1990}).

\bibitem[{\citenamefont{Giamarchi and Lhuillier}(1991)}]{Giamarchi_PRB1991}
\bibinfo{author}{\bibfnamefont{T.}~\bibnamefont{Giamarchi}} \bibnamefont{and}
  \bibinfo{author}{\bibfnamefont{C.}~\bibnamefont{Lhuillier}},
  \bibinfo{journal}{Phys. Rev. B} \textbf{\bibinfo{volume}{43}},
  \bibinfo{pages}{12943} (\bibinfo{year}{1991}).

\bibitem[{\citenamefont{Zhang and Rice}(1988)}]{ZhangRice}
\bibinfo{author}{\bibfnamefont{F.~C.} \bibnamefont{Zhang}} \bibnamefont{and}
  \bibinfo{author}{\bibfnamefont{T.~M.} \bibnamefont{Rice}},
  \bibinfo{journal}{Phys. Rev. B} \textbf{\bibinfo{volume}{37}},
  \bibinfo{pages}{3759} (\bibinfo{year}{1988}).

\bibitem[{\citenamefont{Feiner et~al.}(1996)\citenamefont{Feiner, Jefferson,
  and Raimondi}}]{Raimondi}
\bibinfo{author}{\bibfnamefont{L.~F.} \bibnamefont{Feiner}},
  \bibinfo{author}{\bibfnamefont{J.~H.} \bibnamefont{Jefferson}},
  \bibnamefont{and} \bibinfo{author}{\bibfnamefont{R.}~\bibnamefont{Raimondi}},
  \bibinfo{journal}{Phys. Rev. B} \textbf{\bibinfo{volume}{53}},
  \bibinfo{pages}{8751} (\bibinfo{year}{1996}).

\bibitem[{\citenamefont{M{\"u}ller-Hartmann and Reischl}(2002)}]{Muller}
\bibinfo{author}{\bibfnamefont{E.}~\bibnamefont{M{\"u}ller-Hartmann}}
  \bibnamefont{and} \bibinfo{author}{\bibfnamefont{A.}~\bibnamefont{Reischl}},
  \bibinfo{journal}{Euro. Phys. J. B} \textbf{\bibinfo{volume}{28}},
  \bibinfo{pages}{173} (\bibinfo{year}{2002}).

\bibitem[{\citenamefont{Imada and Miyake}(2010)}]{ImadaMiyake}
\bibinfo{author}{\bibfnamefont{M.}~\bibnamefont{Imada}} \bibnamefont{and}
  \bibinfo{author}{\bibfnamefont{T.}~\bibnamefont{Miyake}},
  \bibinfo{journal}{J. Phys. Soc. Jpn.} \textbf{\bibinfo{volume}{79}},
  \bibinfo{pages}{112001} (\bibinfo{year}{2010}).

\bibitem[{\citenamefont{Wimmer}(2012)}]{PFAPACK}
\bibinfo{author}{\bibfnamefont{M.}~\bibnamefont{Wimmer}}, \bibinfo{journal}{ACM
  Trans. Math. Softw.} \textbf{\bibinfo{volume}{38}}, \bibinfo{pages}{30}
  (\bibinfo{year}{2012}).

\bibitem[{\citenamefont{Watanabe and Imada}(2004)}]{Watanabe}
\bibinfo{author}{\bibfnamefont{S.}~\bibnamefont{Watanabe}} \bibnamefont{and}
  \bibinfo{author}{\bibfnamefont{M.}~\bibnamefont{Imada}}, \bibinfo{journal}{J.
  Phys. Soc. Jpn.} \textbf{\bibinfo{volume}{73}}, \bibinfo{pages}{1251}
  (\bibinfo{year}{2004}).

\bibitem[{\citenamefont{Yokoyama et~al.}(2004)\citenamefont{Yokoyama, Tanaka,
  Ogata, and Tsuchiura}}]{Yokoyama2004}
\bibinfo{author}{\bibfnamefont{H.}~\bibnamefont{Yokoyama}},
  \bibinfo{author}{\bibfnamefont{Y.}~\bibnamefont{Tanaka}},
  \bibinfo{author}{\bibfnamefont{M.}~\bibnamefont{Ogata}}, \bibnamefont{and}
  \bibinfo{author}{\bibfnamefont{H.}~\bibnamefont{Tsuchiura}},
  \bibinfo{journal}{J. Phys. Soc. Jpn.} \textbf{\bibinfo{volume}{73}},
  \bibinfo{pages}{1119} (\bibinfo{year}{2004}).

\bibitem[{\citenamefont{Chang and Zhang}(2010)}]{Chang2010}
\bibinfo{author}{\bibfnamefont{C.-C.} \bibnamefont{Chang}} \bibnamefont{and}
  \bibinfo{author}{\bibfnamefont{S.}~\bibnamefont{Zhang}},
  \bibinfo{journal}{Phys. Rev. Lett.} \textbf{\bibinfo{volume}{104}},
  \bibinfo{pages}{116402} (\bibinfo{year}{2010}).

\bibitem[{\citenamefont{Becca et~al.}(2000)\citenamefont{Becca, Capone, and
  Sorella}}]{Becca}
\bibinfo{author}{\bibfnamefont{F.}~\bibnamefont{Becca}},
  \bibinfo{author}{\bibfnamefont{M.}~\bibnamefont{Capone}}, \bibnamefont{and}
  \bibinfo{author}{\bibfnamefont{S.}~\bibnamefont{Sorella}},
  \bibinfo{journal}{Phys. Rev. B} \textbf{\bibinfo{volume}{62}},
  \bibinfo{pages}{12700} (\bibinfo{year}{2000}).

\bibitem[{\citenamefont{Kashima and Imada}(2001)}]{KashimaImada}
\bibinfo{author}{\bibfnamefont{T.}~\bibnamefont{Kashima}} \bibnamefont{and}
  \bibinfo{author}{\bibfnamefont{M.}~\bibnamefont{Imada}}, \bibinfo{journal}{J.
  Phys. Soc. Jpn.} \textbf{\bibinfo{volume}{70}}, \bibinfo{pages}{2287}
  (\bibinfo{year}{2001}).

\bibitem[{\citenamefont{Morita et~al.}(2002)\citenamefont{Morita, Watanabe, and
  Imada}}]{MoritaImada}
\bibinfo{author}{\bibfnamefont{H.}~\bibnamefont{Morita}},
  \bibinfo{author}{\bibfnamefont{S.}~\bibnamefont{Watanabe}}, \bibnamefont{and}
  \bibinfo{author}{\bibfnamefont{M.}~\bibnamefont{Imada}}, \bibinfo{journal}{J.
  Phys. Soc. Jpn.} \textbf{\bibinfo{volume}{71}}, \bibinfo{pages}{2109}
  (\bibinfo{year}{2002}).

\bibitem[{\citenamefont{Mizusaki and Imada}(2006)}]{MizusakiImada}
\bibinfo{author}{\bibfnamefont{T.}~\bibnamefont{Mizusaki}} \bibnamefont{and}
  \bibinfo{author}{\bibfnamefont{M.}~\bibnamefont{Imada}},
  \bibinfo{journal}{Phys. Rev. B} \textbf{\bibinfo{volume}{74}},
  \bibinfo{pages}{014421} (\bibinfo{year}{2006}).

\bibitem[{\citenamefont{Corboz et~al.}(2008)\citenamefont{Corboz, Troyer,
  Kleine, McCulloch, Schollw\"ock, and Assaad}}]{Corboz}
\bibinfo{author}{\bibfnamefont{P.}~\bibnamefont{Corboz}},
  \bibinfo{author}{\bibfnamefont{M.}~\bibnamefont{Troyer}},
  \bibinfo{author}{\bibfnamefont{A.}~\bibnamefont{Kleine}},
  \bibinfo{author}{\bibfnamefont{I.~P.} \bibnamefont{McCulloch}},
  \bibinfo{author}{\bibfnamefont{U.}~\bibnamefont{Schollw\"ock}},
  \bibnamefont{and} \bibinfo{author}{\bibfnamefont{F.~F.}
  \bibnamefont{Assaad}}, \bibinfo{journal}{Phys. Rev. B}
  \textbf{\bibinfo{volume}{77}}, \bibinfo{pages}{085108}
  (\bibinfo{year}{2008}).

\bibitem[{\citenamefont{Misawa and Imada}(2007)}]{MisawaImada}
\bibinfo{author}{\bibfnamefont{T.}~\bibnamefont{Misawa}} \bibnamefont{and}
  \bibinfo{author}{\bibfnamefont{M.}~\bibnamefont{Imada}},
  \bibinfo{journal}{Phys. Rev. B} \textbf{\bibinfo{volume}{75}},
  \bibinfo{pages}{115121} (\bibinfo{year}{2007}).

\bibitem[{\citenamefont{Moreo et~al.}(1991)\citenamefont{Moreo, Scalapino, and
  Dagotto}}]{Moreo}
\bibinfo{author}{\bibfnamefont{A.}~\bibnamefont{Moreo}},
  \bibinfo{author}{\bibfnamefont{D.}~\bibnamefont{Scalapino}},
  \bibnamefont{and} \bibinfo{author}{\bibfnamefont{E.}~\bibnamefont{Dagotto}},
  \bibinfo{journal}{Phys. Rev. B} \textbf{\bibinfo{volume}{43}},
  \bibinfo{pages}{11442} (\bibinfo{year}{1991}).

\bibitem[{\citenamefont{Plekhanov et~al.}(2003)\citenamefont{Plekhanov,
  Sorella, and Fabrizio}}]{Sorella_VJ}
\bibinfo{author}{\bibfnamefont{E.}~\bibnamefont{Plekhanov}},
  \bibinfo{author}{\bibfnamefont{S.}~\bibnamefont{Sorella}}, \bibnamefont{and}
  \bibinfo{author}{\bibfnamefont{M.}~\bibnamefont{Fabrizio}},
  \bibinfo{journal}{Phys. Rev. Lett.} \textbf{\bibinfo{volume}{90}},
  \bibinfo{pages}{187004} (\bibinfo{year}{2003}).

\bibitem[{\citenamefont{S\'en\'echal et~al.}(2013)\citenamefont{S\'en\'echal,
  Day, Bouliane, and Tremblay}}]{Tremblay_V}
\bibinfo{author}{\bibfnamefont{D.}~\bibnamefont{S\'en\'echal}},
  \bibinfo{author}{\bibfnamefont{A.~G.~R.} \bibnamefont{Day}},
  \bibinfo{author}{\bibfnamefont{V.}~\bibnamefont{Bouliane}}, \bibnamefont{and}
  \bibinfo{author}{\bibfnamefont{A.-M.~S.} \bibnamefont{Tremblay}},
  \bibinfo{journal}{Phys. Rev. B} \textbf{\bibinfo{volume}{87}},
  \bibinfo{pages}{075123} (\bibinfo{year}{2013}).

\bibitem[{\citenamefont{F.~Becca and Sorella}(2009)}]{Becca2009}
\bibinfo{author}{\bibfnamefont{A.~P.} \bibnamefont{F.~Becca},
  \bibfnamefont{L.~Capriotti}} \bibnamefont{and}
  \bibinfo{author}{\bibfnamefont{S.}~\bibnamefont{Sorella}},
  \bibinfo{howpublished}{arXiv:0905.4854} (\bibinfo{year}{2009}).

\bibitem[{\citenamefont{Misawa et~al.}(2012)\citenamefont{Misawa, Nakamura, and
  Imada}}]{misawa2012}
\bibinfo{author}{\bibfnamefont{T.}~\bibnamefont{Misawa}},
  \bibinfo{author}{\bibfnamefont{K.}~\bibnamefont{Nakamura}}, \bibnamefont{and}
  \bibinfo{author}{\bibfnamefont{M.}~\bibnamefont{Imada}},
  \bibinfo{journal}{Phys. Rev. Lett.} \textbf{\bibinfo{volume}{108}},
  \bibinfo{pages}{177007} (\bibinfo{year}{2012}).

\bibitem[{\citenamefont{Kaneko et~al.}(2013)\citenamefont{Kaneko, Morita, and
  Imada}}]{VMC1D}
\bibinfo{author}{\bibfnamefont{R.}~\bibnamefont{Kaneko}},
  \bibinfo{author}{\bibfnamefont{S.}~\bibnamefont{Morita}}, \bibnamefont{and}
  \bibinfo{author}{\bibfnamefont{M.}~\bibnamefont{Imada}},
  \bibinfo{journal}{Journal of Physics: Conference Series}
  \textbf{\bibinfo{volume}{454}}, \bibinfo{pages}{012046}
  (\bibinfo{year}{2013}).

\bibitem[{\citenamefont{Pavarini et~al.}(2001)\citenamefont{Pavarini, Dasgupta,
  Saha-Dasgupta, Jepsen, and Andersen}}]{Andersen}
\bibinfo{author}{\bibfnamefont{E.}~\bibnamefont{Pavarini}},
  \bibinfo{author}{\bibfnamefont{I.}~\bibnamefont{Dasgupta}},
  \bibinfo{author}{\bibfnamefont{T.}~\bibnamefont{Saha-Dasgupta}},
  \bibinfo{author}{\bibfnamefont{O.}~\bibnamefont{Jepsen}}, \bibnamefont{and}
  \bibinfo{author}{\bibfnamefont{O.~K.} \bibnamefont{Andersen}},
  \bibinfo{journal}{Phys. Rev. Lett.} \textbf{\bibinfo{volume}{87}},
  \bibinfo{pages}{047003} (\bibinfo{year}{2001}).

\bibitem[{\citenamefont{Hu et~al.}(2012)\citenamefont{Hu, Becca, and
  Sorella}}]{HuBeccaSorella}
\bibinfo{author}{\bibfnamefont{W.-J.} \bibnamefont{Hu}},
  \bibinfo{author}{\bibfnamefont{F.}~\bibnamefont{Becca}}, \bibnamefont{and}
  \bibinfo{author}{\bibfnamefont{S.}~\bibnamefont{Sorella}},
  \bibinfo{journal}{Phys. Rev. B} \textbf{\bibinfo{volume}{85}},
  \bibinfo{pages}{081110} (\bibinfo{year}{2012}).

\bibitem[{\citenamefont{Sandvik}(1997)}]{Sandvik}
\bibinfo{author}{\bibfnamefont{A.~W.} \bibnamefont{Sandvik}},
  \bibinfo{journal}{Phys. Rev. B} \textbf{\bibinfo{volume}{56}},
  \bibinfo{pages}{11678} (\bibinfo{year}{1997}).

\end{thebibliography}


\end{document}